\documentclass[twocolumn]{aastex631}

\newcommand{\mtwomin}{m_2^{\mathrm{min}}}
\newcommand{\mmin}{m_{\mathrm{min}}}
\newcommand{\mmax}{m_{\mathrm{max}}}
\newcommand{\changed}[1]{#1}
\newcommand{\changedtwo}[1]{#1}

\usepackage{multirow}
\usepackage{printlen}
\usepackage{amsmath,amssymb}
\usepackage{acronym}

\begin{document}
\input{result_macros.sty}

\acrodef{GW}[GW]{gravitational-wave}
\acrodef{BBH}[BBH]{binary black hole}
\acrodef{NSBH}[NSBH]{neutron star--black hole}
\acrodef{AGN}[AGN]{active galactic nuclei}
\acrodef{GWTCthree}[GWTC-3.0]{third Gravitational-Wave Transient Catalog}

\acrodef{CIERA}[CIERA]{Center for Interdisciplinary Exploration and Research in Astrophysics}
\acrodef{STFC}[STFC]{Science and Technology Facilities Council}
\acrodef{NSF}[NSF]{National Science Foundation}

\title{Can Big Black Holes Merge with the Smallest Black Holes?}

\author[0009-0009-9828-3646]{Storm Colloms}
\affiliation{Institute for Gravitational Research, University of Glasgow, Kelvin Building, University Avenue, Glasgow, G12 8QQ, United Kingdom}

\author[0000-0002-2077-4914]{Zoheyr Doctor}
\affiliation{Center for Interdisciplinary Exploration and Research in Astrophysics (CIERA), Northwestern University, 1800 Sherman Avenue, Evanston, IL 60201, USA}

\author[0000-0003-3870-7215]{Christopher P L Berry}
\affiliation{Institute for Gravitational Research, University of Glasgow, Kelvin Building, University Avenue, Glasgow, G12 8QQ, United Kingdom}
\affiliation{Center for Interdisciplinary Exploration and Research in Astrophysics (CIERA), Northwestern University, 1800 Sherman Avenue, Evanston, IL 60201, USA}

\begin{abstract}

Gravitational-wave measurements of the binary black hole population provide insights into the evolution of merging binaries. 
We explore potential correlation between mass and mass ratio with phenomenological population models where the minimum mass of the smaller (secondary) black hole can change with the mass of the bigger (primary) black hole. 
We use binary black hole signals from the third Gravitational-Wave Transient Catalog with and without the relatively extreme mass-ratio GW190814. 
When excluding GW190814, models with a variable minimum mass are disfavoured compared to one with a constant minimum mass, with log Bayes factors of $\ParabolanofourteenuncondBF$ to $\PLnofourteenBF$, indicating that the biggest black holes can merge with the smallest. 
When including GW190814, a parabola model that allows the minimum mass to decrease with increasing primary mass is favoured over a constant minimum-mass model with a log Bayes factor of $\ParabolaBF$. 
When allowing the minimum mass to decrease, the overall population distributions remain similar whether or not GW190814 is included. 
This shows that with added model flexibility, we can reconcile potential outlier observations within our population. 
These investigations motivate further explorations of correlations between mass ratio and component masses in order to understand how evolutionary processes may leave an imprint on these distributions.

\end{abstract}

\keywords{}

\section{Introduction} \label{sec:intro}

\Ac{GW} observations of merging \acp{BBH} from LIGO \citep{LIGOScientific:2014pky}, Virgo \citep{VIRGO:2014yos}, and KAGRA \citep{KAGRA:2020tym} allow us to probe the evolution of \acp{BBH} \citep{Mapelli:2020vfa, Mandel:2018hfr, KAGRA:2021duu}. 
The observed population likely contains contributions from multiple formation channels \citep{zevinOneChannelRule2021, Cheng:2023ddt, Li:2023yyt, Colloms:2025hib}, but the details of the astrophysical evolution of these systems are uncertain \citep{Mandel:2021smh, Belczynski:2021zaz, Mandel:2018hfr}.
Measuring the underlying distribution of \ac{BBH} masses (primary mass $m_1$ and secondary mass $m_2$), mass ratios ($q = m_2/m_1 \leq 1$), spins, redshifts, and the correlations between these parameters from current \ac{GW} observations can inform us of formation mechanisms \citep{Mapelli:2021taw, KAGRA:2021duu, Callister:2024cdx}. 

\ac{BBH} formation processes are expected to leave signatures in the distribution of component masses and mass ratios of merging binaries. 
The mass ratios of binary stars during their evolution depend on the initial fragmentation, accretion during star formation and evolution, and supernovae mechanisms \citep{Bate8134774,Belczynski:2011bn, Moe_2017, neijsselEffectMetallicityspecificStar2019, Zevin:2020gma, Broekgaarden:2022nst, brielUnderstandingHighmassBinary2022}.
Population-synthesis simulations indicate that field binaries that undergo a common-envelope stage can produce \acp{BBH} with mass-ratio distributions with support down to $q\sim0.2$, though this depends on the treatment of the common-envelope process \citep{Zevin:2020gma,Olejak:2021fti}.
Simulations of binaries that undergo stable mass transfer predict this channel cannot produce such unequal-mass binaries, but can produce preferentially more massive black holes than common envelope \citep{vanSon:2021zpk}.
In contrast, \acp{BBH} that merge due to dynamical encounters may have more unequal masses and higher masses than isolated formation channels, especially if they are hierarchical mergers \citep{Rodriguez:2019huv, Sedda:2021vjh, Sedda:2020vwo, Bruel:2023bxl}.
The von Zeipel--Lidov--Kozai effect for field triples can produce \acp{BBH} with mass ratios down to $0.3$ \citep{Martinez:2021tmr}. 
Similarly, \acl{AGN} can produce \acp{BBH} with more unequal masses than isolated binaries \citep{Yang:2019okq, Tagawa:2019osr, Secunda:2020mhd}. 
High-mass \acp{BBH} with unequal masses could therefore be a signature of formation via dynamical encounters or more exotic processes.

Investigations of the mass distributions of black holes have found structure in the primary masses, secondary masses, and the mass ratios of the population. 
\changedtwo{\citet{Fishbach:2019bbm}} found that black holes do not randomly merge with each other regardless of mass; rather \acp{BBH} prefer equal masses.
This is seen in the inferred mass-ratio distribution, which peaks at equal masses when considering a power-law model \citep{LIGOScientific:2020kqk, KAGRA:2021duu}. 
\changed{\citet{Farah:2023swu} modelled the secondary-mass distribution distinctly from the primary-mass distribution, finding that while the distributions of primary and secondary masses are consistent, secondary masses may follow a different distribution than primary masses. 
Different distributions would indicate that primary and secondary black holes undergo different evolutionary processes.}
Works considering more flexible models for the mass-ratio distribution find that the distribution peaks away from equal masses \citep{Sadiq:2023zee, godfreyCosmicCousinsIdentification2023, Rinaldi:2023bbd, Callister:2023tgi, Rinaldi:2025emt}. 
Furthermore, there is evidence that the mass-ratio distribution may change with primary mass \citep{Sadiq:2023zee} and that different features in the primary-mass distribution have different mass-ratio distributions \changedtwo{\citep{Li:2022jge, godfreyCosmicCousinsIdentification2023, Li:2024jzi, Galaudage:2024meo, Roy:2025ktr}}. 
These results indicate that there is information in the mass distributions that could be used to uncover the uncertain astrophysics of \ac{BBH} formation \citep[e.g.,][]{Colloms:2025hib}. 
Exploring parametric model variations for \ac{BBH} masses allows us to test assumptions about the relationship between component masses.
 
\citet{Callister:2024cdx} highlights that the correlations between parameters, in addition to the parameters' marginal distributions, can be informative. 
When investigating correlations, it is necessary to check whether results are accurate descriptions of the population, or whether they are driven by the enforced model constraints.
\citet{Adamcewicz:2022hce} investigated the anti-correlation between mass ratio and effective inspiral spin.
By fixing the shape of the marginal distributions, they avoid the case where the correlation results from an improved fit of the marginal distributions. 
The same can be true in the other direction: inferences of marginal distributions could be driven by assumed correlations (or the lack thereof) imposed by the chosen model. 
Potentially, individual source constraints that are not well accounted for by model correlations could influence the measured marginal distributions, especially if a correlation must be distorted to fit them. 
Updating population models so that outliers are well accounted for and correlations are appropriately incorporated helps us understand the entire population of \acp{BBH}.

Modelling \ac{BBH} mass distributions requires a minimum black hole mass.
This raises the question of the classification of merging objects with a mass that could correspond either to a black hole or a neutron star \citep{Farah:2021qom}. 
It is unknown whether the maximum neutron star mass is the same as the minimum black hole mass. 
The maximum neutron star mass is uncertain, and depends on the neutron star equation of state \citep{Rutherford:2024srk, Koehn:2024set}. 
Meanwhile, uncertainties in the minimum black hole mass come from the details of supernovae and core collapse \citep{Fryer:2011cx, Liu:2020uba, Patton:2021gwh,Fryer:2022lla}.
Following X-ray binary observations, it has been proposed that the maximum neutron star mass and the minimum black hole mass are not equal~\citep{Bailyn:1997xt,Ozel:2010su,Farr:2010tu,Siegel:2022gwc}, and there is a mass gap with a dearth of black holes.
Observations of objects with masses $\sim 2.5$--$5~{M}_{\odot}$ indicate that at least some of the proposed gap is populated, though the rate of these objects is uncertain \citep{Thompson:2018ycv, LIGOScientific:2020zkf, Jayasinghe:2021uqb, LIGOScientific:2021djp, LIGOScientific:2024elc, Fishbach:2025bjh}. 
Including low-mass black holes from \ac{GW} observations can allow us to measure the minimum black hole mass \citep{Ray:2025aqr}.

Observations of (potential) low-$m_2$ \acp{BBH} include GW190814, which has $q=0.112^{+0.008}_{-0.009}$ and $m_2=2.6^{+0.1}_{-0.1} {M}_{\odot}$ \citep{LIGOScientific:2020zkf,LIGOScientific:2021usb}.
Understanding the nature of GW190814's source as a \ac{BBH} or a \acl{NSBH} requires understanding if it is an outlier to the \ac{BBH} population, and whether its outlier status is because of population misspecification.
While \citet{LIGOScientific:2020kqk} and \citet{Essick:2021vlx} find GW190814 to be an outlier, \citet{Farah:2021qom} consider models where it is not by including the full population of neutron stars and black holes, and considering a partially full mass gap. 
The outlier status of observations thus depends on the choice of population model.
We seek to understand how to improve our models and incorporate observations that are currently challenging to fit.

Current models for the \acp{BBH}' mass distribution assume that the minimum black hole mass is constant with the primary black hole mass. 
Given the preference for equal-mass mergers and the structure in the distribution of \ac{BBH} masses, this may not be the case. 
We investigate if there is evidence for an evolving minimum secondary mass with primary mass: can black holes with higher masses merge with the lowest-mass black holes? 
We introduce models for the mass-ratio distribution of \acp{BBH} specifying a minimum secondary black hole mass that evolves with the primary mass (Section~\ref{sec:methods}). 
Using these models on observations from the \acl{GWTCthree} \citep[\acsu{GWTCthree};][]{LIGOScientific:2021djp} we infer the primary-mass and mass-ratio distributions (Section~\ref{sec:results}). 
Our conclusions for the inferred evolution of minimum secondary mass, and measured primary-mass and mass-ratio distributions are discussed in Section~\ref{sec:conc}.

\section{Methods}\label{sec:methods}

We estimate a set of hyperparameters $\Lambda$ that govern the shape of the mass, spin and redshift distributions of \ac{GW} observations from \ac{GWTCthree} \citep{LIGOScientific:2021djp} using a hierarchical Bayesian inference \citep{mandelExtractingDistributionParameters2019, thraneIntroductionBayesianInference2019, vitaleInferringPropertiesPopulation2021}. 
For each population distribution we choose a parametric model. 
We assume the distribution of the primary mass, spin and redshift distributions to be the default parametric models used in \cite{KAGRA:2021duu}, and introduce new models for the mass-ratio distribution as described in Section~\ref{sec:models}.
Our inference framework is described in more detail in section Section~\ref{sec:statframe} and the observations we use are described in Section~\ref{sec:obs}.

\subsection{Variable minimum-mass models}\label{sec:models}

To investigate the correlation between the minimum mass and the primary mass, we model the mass-ratio distribution of \acp{BBH} with a minimum secondary mass that varies with the primary mass. 
We include this as a modification to a power-law model of mass ratio \citep{fishbachWhereAreLIGO2017, Kovetz:2016kpi}, such that the mass-ratio distribution $p(q\mid m_1)$ with a power-law index $\beta$ and a minimum secondary mass $\mtwomin(m_1)$ follows
\begin{equation}
    \begin{split}
    p(q\mid m_1) &= \left(1 + \beta\right)\left[1 - \frac{\mtwomin(m_1)}{m_1}\right]^{-(1 + \beta)} q^\beta \\
     &\propto q^{\beta} \qquad \qquad \left\{\frac{\mtwomin(m_1)}{m_1} \leq q \leq 1\right\}.
    \end{split}
\end{equation}
This allows a varying lower bound on the secondary mass $m_2$ such that $m_2 \geq \mtwomin(m_1)$, the smallest possible secondary mass, and therefore the minimum mass ratio varies with the primary mass.

We consider two forms for how the minimum secondary mass varies with primary mass.
Neither has significant astrophysical motivation, but both are simple progressions from a fixed minimum mass that enable us to explore whether the data support more complicated models. 
The two models complement each other in their analytic forms, and minimise additional model hyperparameters.

First, we consider a \textsc{Power Law} form that adds one additional hyperparameter:
\begin{equation}
\label{eq:PLm2min}
    \mtwomin= \mmin + (\mmax - \mmin)\left(\frac{m_1-\mmin}{\mmax - \mmin}\right)^\gamma,
\end{equation}
where $\gamma \geq 1$, $\mmin$ is the smallest primary black hole mass, and $\mmax$ is the maximum (stellar-mass) black hole mass. 
The \textsc{Power Law} model is a strictly increasing function that sets $\mmin$ as the minimum black hole mass for both the primary and secondary objects, and  $\mtwomin(\mmax)=\mmax$. 
As $\gamma$ increases, lower $\mtwomin$ is allowed for each $m_1$. 
Inspired by observations of binaries preferring more equal-mass components, this model prevents high-mass primaries from merging with low-mass secondaries.

Second, we consider a \textsc{Parabola} model for $\mtwomin$ that adds two additional hyperparameters:
\begin{equation}
\label{eq:parabolam2min}
    \mtwomin(m_1) = \mmin + \xi(m_1 - \mmin) + \zeta(m_1 - \mmin)^2.
\end{equation}
This functional form automatically enforces that $\mtwomin(\mmin) = \mmin$. 
We apply additional constraints to ensure that the model remains physical. 
We require that $\mtwomin(m_1)< m_1$ for all $m_1 < \mmax$. 
Equivalently,
\begin{align}
        \mmax > \mmin + \xi(\mmax - \mmin) \nonumber \\
        + \left. \zeta(\mmax - \mmin)^2 . \right.
\end{align}
This yields the bound
\begin{equation}
    \zeta < \frac{1-\xi}{\mmax - \mmin}.
    \label{eq:m1-bound}
\end{equation}
We consider two variations of this model.

The \textsc{Increasing Parabola} model has $\xi > 0$ and $\zeta > 0$ such that $\mtwomin$ is always increasing with $m_1$.

The \textsc{Relaxed Parabola} model allows for a decreasing $\mtwomin$ with $m_1$ by allowing for negative $\xi$. 
This means that the minimum secondary mass can be less than $\mmin$.
For this case, we require explicitly enforcing that $\mtwomin(m_1)>0$ for all $m_1$. 
This necessitates that Eq.~\eqref{eq:parabolam2min} has no real roots, and
\begin{equation}
    \zeta > \frac{\xi^2}{4 \mmin}.
    \label{eq:increase-bound}
\end{equation}
Combining Eq.~\eqref{eq:m1-bound} and Eq.~\eqref{eq:increase-bound} sets a  limit on $\xi$ such that
\begin{align}
        \frac{1-\xi}{\mmax - \mmin} > \frac{\xi^2}{4 \mmin} , 
\end{align}
and hence,
\begin{align}
        -2 \frac{\sqrt{\mmin}}{\sqrt{\mmax}-\sqrt{\mmin}} < \xi < 2 \frac{\sqrt{\mmin}}{\sqrt{\mmax}+\sqrt{\mmin}}.
\end{align}
We consider these constraints in our priors on $\xi$ and $\zeta$, as detailed in Table~\ref{tab:mminmodelpriors}. 

\begin{figure}
    \centering
    \includegraphics[width=0.9\linewidth]{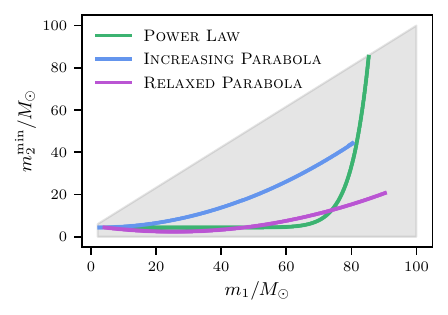}
    \caption{\changed{Example evolution of $\mtwomin$ with $m_1$ for the \textsc{Power Law}, \textsc{Increasing Parabola}, and \textsc{Relaxed Parabola} model variations we consider, for some fiducial hyperparameters within our prior range. The shaded region indicates where $m_1>\mtwomin>0$.}}
\label{fig:modelreps}
\end{figure}

\changed{Figure \ref{fig:modelreps} shows graphical representations of each model for some fiducial hyperparameters.}
\begin{table*}[]
    \centering
    \begin{tabular}{lcc}
    \hline
    Model 
    & {Hyperparameter priors} 
    & {Minimum secondary mass}\\
    \hline\hline
     & & \\
      \textsc{Power Law} & $\gamma \sim \mathcal{U}(1, 35)$  & Eq.~\eqref{eq:PLm2min}  \\
     &  & \\ \hline
     & & \\
      \textsc{Increasing Parabola} & $\xi \sim \mathcal{U}(0, 1)$  & Eq.~\eqref{eq:parabolam2min}  \\
      & $\displaystyle \zeta \sim\mathcal{U} \left(0,\frac{1-\xi}{\mmax - \mmin} \right)$ &  \\
     &  & \\ \hline
     & & \\
     \textsc{Relaxed Parabola} & 
        $\displaystyle \xi \sim \mathcal{U}\left(-2\frac{\sqrt{\mmin}}{\sqrt{\mmax}-\sqrt{\mmin}}, 2 \frac{\sqrt{\mmin}}{\sqrt{\mmax}+\sqrt{\mmin}}\right)$  & Eq.~\eqref{eq:parabolam2min} \\ 
     & $\displaystyle \zeta \sim \mathcal{U}\left(\frac{\xi^2}{4 \mmin} ,\frac{1-\xi}{\mmax - \mmin} \right)$ &  \\
     &  &  \\ \hline
    \end{tabular}
    \label{tab:mminmodelpriors}
    \caption{
        Prior ranges on $\gamma$, $\xi$ and $\zeta$ hyperparameters for the three minimum secondary-mass models we consider, \textsc{Power Law}, \textsc{Increasing Parabola},  and \textsc{Relaxed Parabola}, and their corresponding forms for the minimum secondary mass.}
\end{table*}

\subsection{Statistical framework}
\label{sec:statframe}

To investigate the overall distribution of the \ac{BBH} population, we jointly infer the distribution of \ac{BBH} primary masses, mass ratios, spin magnitudes and tilts, and redshifts.
We model the distribution of \ac{BBH} primary masses, mass ratios, spin magnitudes $\chi_{1,2}$ and tilts $\theta_{1,2}$, and redshifts $z$ as a separable probability distribution conditional on population hyperparameters $\Lambda$,
\begin{equation}
    \begin{gathered}\label{eq:pm1qz}
    p(m_1,q,\chi_{1,2},\cos\theta_{1,2},z\mid\Lambda) = \left. p(m_1\mid\Lambda) \times p(q|m_1,\Lambda) \right. \times\\
    p(\chi_{1,2}\mid\Lambda)\times p(\cos\theta_{1,2}\mid\Lambda)\times p(z\mid\Lambda),
    \end{gathered}
\end{equation}
where  the mass-ratio distribution includes the \textsc{Power Law}, \textsc{Increasing Parabola} or \textsc{Relaxed Parabola} minimum secondary-mass model described above, and the other observables have distributions that follow the default parametric models used in \citet{KAGRA:2021duu}.
We use the \textsc{Power Law+Peak} model for $p(m_1)$   \citep{Talbot:2018cva}.
This models the $m_1$ distribution as a two-component mixture of a power law  of slope $\alpha$ and a Gaussian component.
We choose similar priors on the default model parameters as \citet{KAGRA:2021duu}, with some refinements. 
We use primary-mass power-law index $\alpha \sim \mathcal{U}(-4, 12)$, mass-ratio power-law index $\beta \sim \mathcal{U}(-2, 7)$, primary minimum mass $m_\mathrm{min} \sim \mathcal{U}(2, 6)\,M_\odot$, and maximum mass $m_\mathrm{max} \sim \mathcal{U}(62, 100)\,M_\odot$.
Both spin magnitudes are modelled with an identical beta distribution \changed{\citep{Wysocki:2018mpo}}, and the spin tilts a mixture of isotropic and normally distributed spin tilts \changed{\citep{Talbot:2017yur}}.
The redshift distribution $p(z)$ is modelled with a power law \citep{Fishbach:2018edt} of index $\kappa \sim \mathcal{U}(-2, 8)$ .

To infer hyperposteriors on $\Lambda$  we use the population-inference package \texttt{gwpopulation\_pipe} \changed{\citep{Talbot:2024yqw, gwpop_pipe}} with the nested sampler \texttt{dynesty} \citep{Speagle_2020}. 
The inference imposes a cut on the likelihood variance of $1$ during sampling \citep{Talbot:2023pex}.

\subsection{Gravitational-wave observations}
\label{sec:obs}
We use the observations from \ac{GWTCthree} \citep[][]{LIGOScientific:2021djp} selected with a false alarm rate threshold of less than $1~\mathrm{yr}^{-1}$.
This is identical to the set of observations used in \citet{KAGRA:2021duu}.
We consider only signals with  \ac{BBH} sources and the ambiguous GW190814 \citep{LIGOScientific:2020zkf, LIGOScientific:2021usb}.
As GW190814's secondary object may be above the expected maximum mass for a neutron star, we compare the results with and without GW190814, and investigate its potential as an outlier to the \ac{BBH} population. 
GW200210\_092254~\citep{LIGOScientific:2021djp}, which has a similar secondary mass and mass ratio to GW190814, is not included in our analysis as it does not satisfy the false alarm rate threshold.

We use a combined injection set of semi-analytic injections at the sensitivity of the first and second observing runs, and real injections at the third observing run sensitivity to account for selection effects \changed{\citep{KAGRA:2021duu, GWTC3_sensitivity}}. 
We consider the real injections with a false alarm rate of less than $1~\mathrm{yr}^{-1}$ detectable, and the semi-analytic injections we cut in network signal-to-noise ratio of $10$ \citep{LVKsensitivity2018}.

\section{Results} \label{sec:results}

\subsection{Evolution of minimum secondary mass} \label{subsec:m2minresults}
\begin{figure*}
\centering
\includegraphics[width=0.94\textwidth]{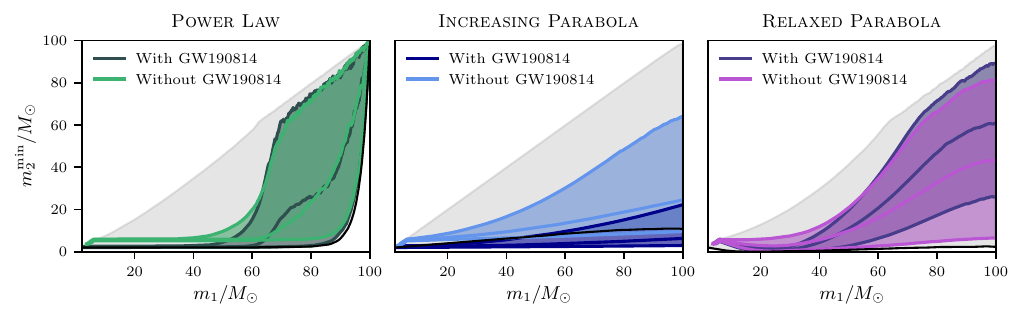}
\caption{Inferred evolution of the minimum secondary mass with primary mass for the \textsc{Power Law} (left), \textsc{Increasing Parabola} (middle), and \textsc{Relaxed Parabola} (right) models. 
Each panel shows the resulting median and $90\%$ credible interval on the minimum secondary mass, including and excluding GW190814. 
The grey shows the $99\%$ credible interval for the prior for each model, with the lower edge in black. 
The interval is evaluated over $m_1\in[2,100]~M_{\odot}$, where the plotted range extends to $m_1>m_{\mathrm{max}}$ for some individual hyperposterior samples. 
With GW190814, the entire distribution of $\mtwomin$ is lower, apart from the \textsc{Relaxed Parabola} model, which decreases to allow for the lower secondary mass at $m_1\sim20\,M_{\odot}$.
Due to the driving of the prior, $\mtwomin$ then increases at higher $m_1$ for this model.
Otherwise, there is no strong evidence for an increasing $\mtwomin$ with $m_1$ when accounting for model limitations.
\label{fig:mtowminmone}}
\end{figure*}
We show the inferred evolution of $\mtwomin$ with $m_1$ in Figure~\ref{fig:mtowminmone}, with and without GW190814, for our three model variations. 
Uncertainties grow at larger $m_1$ where we have fewer observations.

\begin{figure}
\centering
\includegraphics[width=\columnwidth]{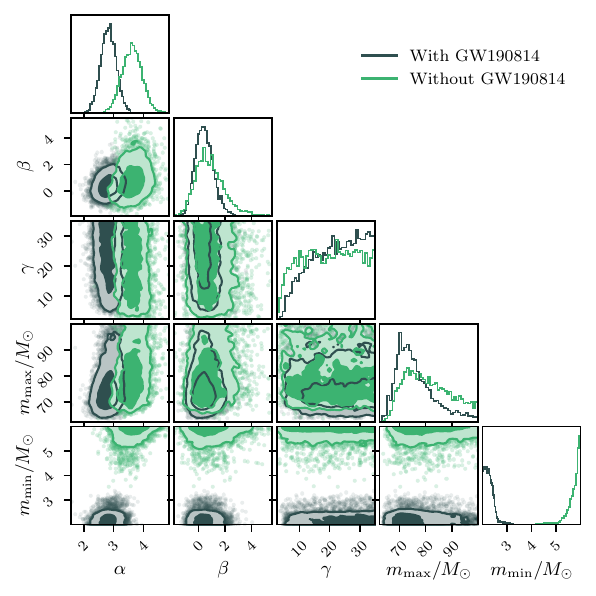}
\caption{Hyperposteriors on the key mass and mass-ratio hyperparameters with \textsc{Power Law} model, with and without GW190814.
The $\mtwomin$ hyperparameter $\gamma$ prefers high values, supporting a non-increasing or shallowly increasing $\mtwomin$ with $m_1$. 
The other hyperparameters show disagreement between the choice to include or exclude GW190814.
\label{fig:PLcorner}}
\end{figure}

Under the \textsc{Power Law} minimum secondary-mass model, the evolution of $\mtwomin$ is preferentially flat until high primary masses, where the uncertainty on $\mtwomin$ grows. 
At high $m_1$, $\mtwomin$ is forced to increase to $m_{\mathrm{max}}$ because of the functional form of the model. 
The hyperposteriors in Figure~$\ref{fig:PLcorner}$ show that high $\gamma$ is preferred, meaning that the evolution of $\mtwomin$ is flatter with $m_1$, before sharply increasing as $m_1$ approaches $\mmax$. 
The results are restricted by the prior on $\gamma$, as we allow a maximum $\gamma$ of $35$, while a flat evolution in $\mtwomin$ would require $\gamma\rightarrow\infty$.
With our prior choice, we find that $\gamma>\PLgammalowlim$ with GW190814 and $\gamma>\PLgammalowlimnofourteen$ without GW190814 at $90\%$ credibility. 
The preference for a flatter evolution of $\mtwomin$ means that there is little support for a higher minimum secondary mass at higher primary masses. 
While this behaviour is consistent whether or not GW190814 is included, the overall minimum mass $\mmin$ and the primary-mass power-law index $\alpha$ are different when including GW190814.
Without GW190814, the minimum mass is higher, with $\mmin =\PLmminnofourteen~M_{\odot}$, while with GW190814, $\mmin=\PLmmin~M_{\odot}$ to accommodate GW190814's secondary mass.

\begin{figure}
\centering
\includegraphics[width=\columnwidth]{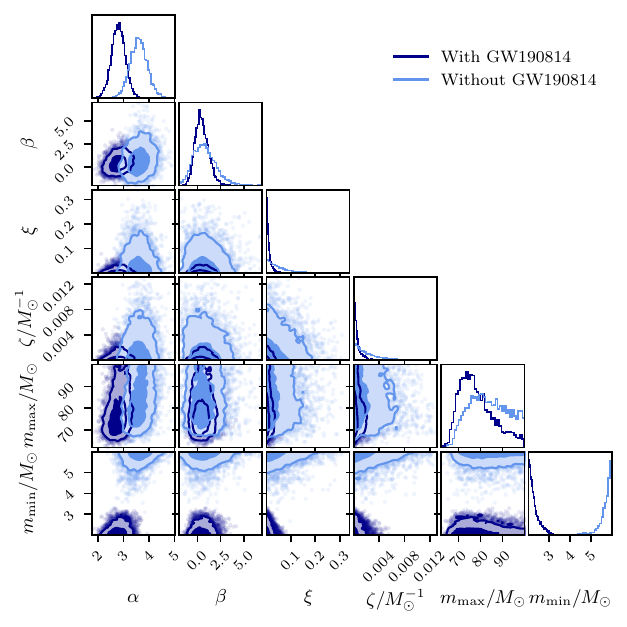}
\caption{Hyperposteriors on the key mass and mass-ratio hyperparameters with the \textsc{Increasing Parabola} model, with and without GW190814.
The $\mtwomin$ hyperparameters, $\xi$ and $\zeta$ are railing at $0$, showing a preference for a non-increasing $\mtwomin$ with $m_1$. 
The other hyperparameters show disagreement between the choice to include or exclude GW190814.
\label{fig:parabolacorner}}
\end{figure}

For the \textsc{Increasing Parabola} model, the inferred $\mtwomin$ behaviour is similar to the \textsc{Power Law} result, where $\mtwomin$ is consistent with being flat with $m_1$ for both sets of results. 
Hence, there is no evidence that the minimum secondary mass changes with primary mass when considering this model. 
This can be seen in the hyperposteriors on $\xi$ and  $\zeta$, as plotted in Figure~\ref{fig:parabolacorner}, with $\xi$ and $\zeta$ both peaking at $0$. 
Similarly to the \textsc{Power Law} model, when including GW190814, $\mmin$ is forced to be lower to include its secondary mass, and $\alpha$ is smaller. 
At large primary masses, the uncertainty on the minimum secondary mass is high, though this is reduced when including GW190814.
This may result from the restriction that $\mtwomin$ cannot increase above $\sim2.6~M_\odot$ at $m_1\sim20~M_\odot$, as can be seen by the more tightly constrained $\xi$ and $\zeta$ when including GW190814.

\begin{figure}
\centering
\includegraphics[width=\columnwidth]{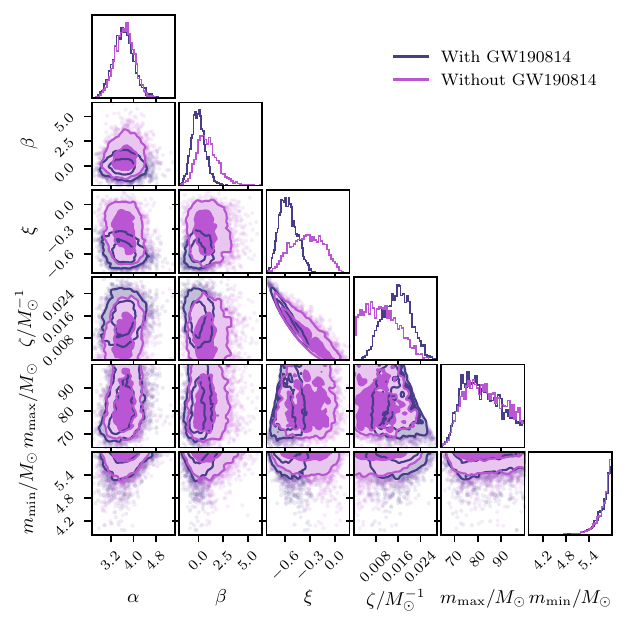}
\caption{Hyperposteriors on the key mass and mass-ratio hyperparameters with the \textsc{Relaxed Parabola} model, with and without GW190814.
The $\mtwomin$ hyperparameters, $\xi$ and $\zeta$ are consistent with zero without GW190814, but negative $\xi$ is preferred when including GW190814. 
The other hyperparameters are in agreement between the two choices of observations.
\label{fig:paraboladecreasingcorner}}
\end{figure}

In the \textsc{Relaxed Parabola} model, $\mtwomin$ remains largely unconstrained at high $m_1$, while $\mtwomin(\mmax)=\mmin$ is not ruled out when not including GW190814.
With GW190814, $\mtwomin$ decreases below $\mmin$ until $m_1\sim20~M_\odot$, then increases at high $m_1$, preferring $\mtwomin > \mmin$ at higher primary masses.
The dip in $\mtwomin$ at $m_1\sim20~M_\odot$ accommodates GW190814's secondary mass, while the increase of $\mtwomin$ at high $m_1$ is driven by the prior constraints when including negative $\xi$.
For both sets of observations, $\mmin$ does not significantly change as $\mtwomin$ is instead allowed to decrease to allow for the secondary mass of GW190814. 
This is because $\mmin$ is no longer describing the lowest secondary mass of the observations, but the lowest primary mass.
As shown in Figure~\ref{fig:paraboladecreasingcorner}, while the hyperposteriors on $\xi$ and $\zeta$ are different with the different sets of observations, the other hyperposteriors show less variation when including GW190814 than results with the strictly increasing $\mtwomin$ models.
This hints that GW190814 is no longer an outlier under the model assumptions that allow $\mtwomin$ to decrease with $m_1$.

We compare each of our analyses to the results with the default \textsc{Power Law + Peak} model from \cite{KAGRA:2021duu}, with $\mtwomin$ fixed to be equal to $\mmin$.
We find that our models are generally not preferred to the default model without a variable $\mtwomin$, as shown in Table~\ref{tab:BFs}. 
This reinforces the result that there is no evidence that $\mtwomin$ (strictly) increases with $m_1$.
The exception is the \textsc{Relaxed Parabola} model, which is preferred with a log Bayes factor of $\log_{10}\mathcal{B}=\ParabolaBF$ when including GW190814.
Therefore, if GW190814's source is a \ac{BBH}, this indicates that there is structure in the mass-ratio and component-mass distributions that should be accounted for by population models.
\begin{table}[]
\centering
\begin{tabular}{lcc}
\hline
 & \multicolumn{2}{c}{{$\log_{10}\mathcal{B}$}} \\ \cline{2-3}
{}   & With & Without \\ 
{Model}   & GW190814 & GW190814 \\ \hline\hline
\textsc{Power Law} & $\PLBF$ & $\PLnofourteenBF$ \\
\textsc{Increasing Parabola} & $\ParabolauncondBF$  & $\ParabolanofourteenuncondBF$ \\ 
\textsc{Relaxed Parabola} & $\hphantom{-}\ParabolaBF$ & $\ParabolanofourteenBF$ \\  \hline
\end{tabular}
\caption{Log Bayes factors between the \textsc{Power Law}, \textsc{Increasing Parabola}, and \textsc{Relaxed Parabola} minimum secondary-mass models and the default model without evolution of the minimum secondary mass from \cite{KAGRA:2021duu}. 
All models allowing for an evolving minimum secondary mass with primary mass are disfavoured compared to the default model, with the exception of the \textsc{Relaxed Parabola} model when including GW190814.}
\label{tab:BFs}
\end{table}

\subsection{Mass-ratio and primary-mass distributions} 
\label{subsec:PLmodel}

\begin{figure*}
\centering
\includegraphics[width=0.92\textwidth]{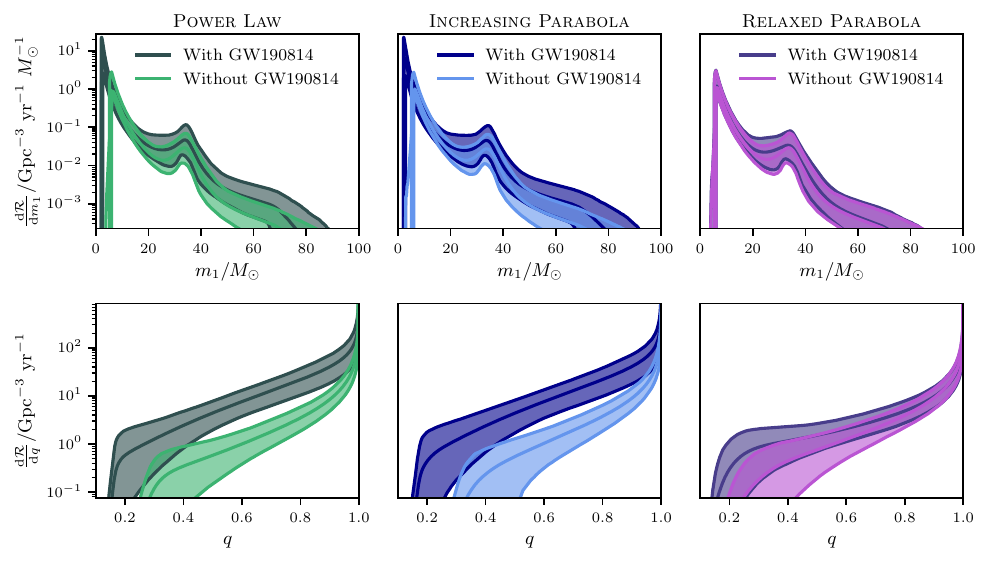}
\caption{The inferred primary-mass (top) and mass-ratio (bottom) distributions for the \textsc{Power Law} (left), \textsc{Increasing Parabola} (middle), and \textsc{Relaxed Parabola} (right) models, each with and without GW190814. The solid line shows the posterior population distribution and the shaded region is the $90\%$ credible interval for each distribution.
For the \textsc{Increasing Parabola} and \textsc{Power Law} models, the distributions are different when including GW190814, while for the \textsc{Relaxed Parabola} model the intervals are consistent.
\label{fig:ppds}}
\end{figure*}

To understand the impact of a variable $\mtwomin$ on the mass-ratio distribution, we show $p(m_1)$ and $p(q)$ for our three model variations in Figure~\ref{fig:ppds}.
For the \textsc{Power Law} and \textsc{Increasing Parabola} models, neither $p(m_1)$ nor $p(q)$ agree between the results with and without GW190814. 
Both of these models find a different $\mmin$ for the different sets of observations, resulting in a different $p(m_1)$ distribution at lower masses. 
Without GW190814, these models also are not required to go to as low mass ratios. 
With these strictly increasing $\mtwomin$ models, inclusion of GW190814 changes the overall primary-mass distribution despite its $m_1$ being unexceptional. 
For the \textsc{Relaxed Parabola} model, there is more agreement between both primary-mass and mass-ratio distributions. 
While $p(q)$ does extend to lower values with GW190814, this is expected due to its low mass ratio.
The added flexibility in the mass-ratio model thus allows for GW190814-like systems to fit with the bulk \acp{BBH} population, and they do not need to be treated as population outliers.

\begin{figure*}
\centering
\includegraphics[width=0.92\textwidth]{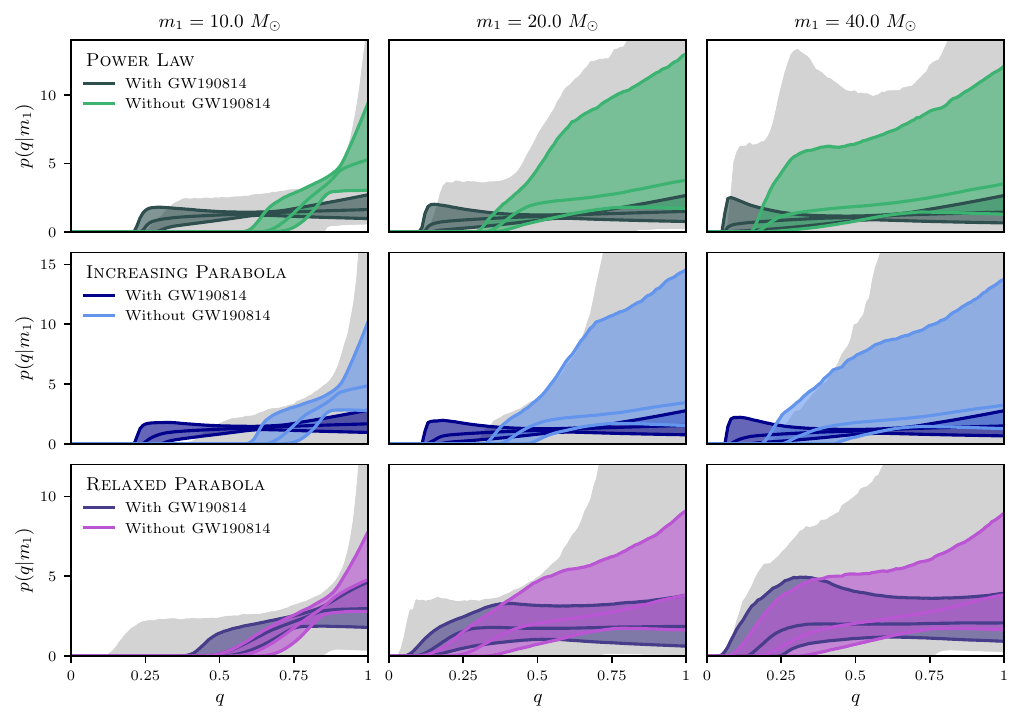}
\caption{Slices of the median and $90\%$ interval for the mass-ratio distribution at primary masses of $10~M_\odot$ (left), $20~M_\odot$ (middle), and $40~M_\odot$ (right) for the \textsc{Power Law} (top), \textsc{Increasing Parabola} (middle), and \textsc{Relaxed Parabola} (bottom) models, each with and without GW190814.
\changed{The $99\%$ credible interval of the prior at each slice is shown in grey.}
For the \textsc{Increasing Parabola} and \textsc{Power Law} models, the distributions are different when including GW190814, while for the \textsc{Relaxed Parabola} model the intervals are more consistent.
\label{fig:pq}}
\end{figure*}

Figure~\ref{fig:pq} shows $p(q)$ at slices of $m_1$ for the three model variations.
This shows the strictly increasing $\mtwomin$ models' inconsistency between the mass-ratio distributions for the results with and without GW190814, especially for lower primary masses. 
At $m_1=10~M_\odot$ for the \textsc{Power Law} and \textsc{Increasing Parabola} models, the mass-ratio distribution supports much lower mass ratios when including GW190814 because of the overall lower $\mmin$. 
As $\mmin\sim5~M_\odot$ without GW190814, the lowest possible mass ratio is $q\sim0.5$ at $m_1=10~M_\odot$, while when including GW190814, $\mmin\sim2~M_\odot$, making the lowest possible mass ratio $q\sim0.2$. 
Extreme mass ratios at $m_1=10~M_\odot$ are not well supported by the prior, and GW190814 is highly informative for the inference. 
At all mass slices, $p(q)$ is flatter for inference with GW190814, driven by the available model space. 
The difference at $m_1=10~M_\odot$ is reconciled for the \textsc{Relaxed Parabola} model, as $\mmin$ is in agreement whether or not GW190814 is included. 

Compared to our analyses, \citet{Callister:2023tgi} and \citet{Sadiq:2023zee} find that $p(q)$ has more support for unequal-mass mergers when considering data-driven models. 
These models do not consider a parametrised $\mmin$ and therefore have more flexibility to extend support to lower $q$ even at low $m_1$. 
\citet{Sadiq:2023zee}, when excluding GW190814, find that $p(q\mid m_1=10~M_{\odot})$ has support down to $q \sim 0.3$, while the distribution peaks away from equal masses at higher $m_1$.
\changedtwo{Using parametric models, \citet{Li:2022jge} find that higher-mass \acp{BBH} are more likely to be equal mass than lower-mass binaries when excluding GW190814.
Meanwhile, \citet{Galaudage:2024meo}, \citet{Li:2024jzi}, and \citet{Roy:2025ktr} find that \acp{BBH} around $m_1\sim35~M_\odot$ are more likely to prefer equal masses. 
As shown in Figure~\ref{fig:pq}, our models induce that the mass-ratio distribution varies with primary mass when excluding GW190814: at the lowest masses, the distribution looks steep because the range of potential mass ratios is tightly constrained by the minimum mass, while at larger masses the preferred distribution is comparatively flatter, though with broad uncertainties. 
The variation between results across the literature illustrates how different model constraints and prior assumptions can influence the interpretation of results, and highlights the importance of exploring alternative choices.
}

\section{Conclusions} \label{sec:conc}

We investigated the correlation between the mass ratio and primary mass of merging \acp{BBH} to understand whether the biggest stellar-mass black holes can merge with the smallest. 
We introduced population models that included a minimum secondary mass $\mtwomin(m_1)$ that varies with primary mass. 
We considered three model variations for $\mtwomin$, which we used with a power-law model for the mass-ratio distribution and a \textsc{Power Law + Peak} distribution for primary mass \citep{Talbot:2018cva, KAGRA:2021duu}.
Using these models we inferred how the minimum secondary mass evolves for \ac{GWTCthree} observations  \citep{LIGOScientific:2021djp}.

Our results show that there is currently no evidence that $\mtwomin$ increases with $m_1$ when excluding GW190814. 
We find log Bayes factors of $\log_{10}\mathcal{B}=\ParabolanofourteenuncondBF$ to $\PLnofourteenBF$ for our models compared to a constant minimum-mass model. 
We repeated the same analysis including GW190814. 
If we enforce that $\mtwomin$ is strictly increasing with $m_1$, we find that the overall minimum black hole mass has to be lower to account for GW190814's secondary mass, while the primary-mass and mass-ratio distributions change with inclusion of GW190814. 
This suggests that GW190814 should be considered as an outlier for these population models. 
With a model allowing $\mtwomin$ to decrease with $m_1$, $\mtwomin$ dips to allow GW190814's secondary mass, and the minimum primary mass is in agreement with the inference excluding GW190814.
The added flexibility in $\mtwomin$ means that under this model the differences in $p(m_1)$ and $p(q)$ are reconciled when doing a leave-one-out analysis, no longer causing GW190814 to be an outlier.
This model is favoured when including GW190814 in the population compared to a constant minimum-mass model, with $\log_{10}\mathcal{B}=\ParabolaBF$.
Hence, there is support for an evolving $\mtwomin$ when including GW190814's source in the \ac{BBH} population.

The preference for a near-constant $\mtwomin$ with $m_1$ means that we do not rule out the possibility of high-primary mass, relatively extreme mass-ratio \acp{BBH}. 
However, our model uncertainties are greatest at large primary masses due to a combination of a lack of observations and model limitations. 
    
While our models provide one way to understand correlations between primary mass and mass ratio, improvements to our model could be made to ensure robustness and avoid model misspecification. 
Our model that allows for decreasing $\mtwomin$ enforces that the minimum secondary mass is always positive, but currently does not place other bounds on $\mtwomin$, e.g., it could go to subsolar masses. 
Future work could update this assumption by imposing a non-zero lower bound on $\mtwomin$, representing a minimum black hole mass, which may help us understand the classification of GW190814. 
Additional model flexibility also could be added, for example by relaxing the assumption that $\mtwomin(\mmin) = \mmin$; 
\citet{Farah:2023swu} find that the inferred minimum primary and minimum secondary masses of the \ac{BBH} population are in agreement within current uncertainties, but did not include GW190814 in their analysis. 
Our parametric forms for the minimum secondary mass are also not motivated by particular astrophysics or trend in the data. 
A data-driven description of the primary-mass and mass-ratio distributions may give more insight into the underlying correlations, unrestricted by the choice of parametric form.
Furthermore, we specify a power-law distribution in mass ratio, restricting the mass-ratio distribution to be strictly monotonic. 
However, models with more relaxed assumptions about the shape of the distribution show support for the mass-ratio distribution to peak away from $1$ \citep{Sadiq:2023zee, godfreyCosmicCousinsIdentification2023, Rinaldi:2023bbd, Callister:2023tgi, Rinaldi:2025emt}. 
\changedtwo{Other works find that the mass-ratio distribution changes with primary mass, either continuously or for different subpopulations, motivating the exploration of a variable shape of the distribution in addition to a variable lower bound \citep{Li:2022jge, Baibhav:2022qxm, Galaudage:2024meo, Roy:2025ktr}.}
Considering a mass-ratio distribution that more accurately reflects the data would help avoid biases driven by an inaccurate models.

Other observable parameters such as the spin or eccentricity could provide more information about the astrophysical nature of such \acp{BBH}.  
Works looking at the correlations between mass and spin may identify subpopulations of sources from similar evolutionary pathways \citep[e.g.,][]{Safarzadeh:2020mlb,Heinzel:2023hlb,Ray:2024hos,Antonini:2024het}. 
If the mass-ratio distribution is jointly correlated with component mass and spin distributions, neglecting to model these correlations could lead to biased inference \citep{Alvarez-Lopez:2025ltt}.
Jointly investigating mass, mass-ratio, and spin distributions could help us understand how these parameters are correlated, and therefore give more insight into the astrophysical evolution of the population. 
However, adding additional parameters to our models and investigating the joint correlations between mass, mass ratio, and spin, requires having sufficient data to constrain the model parameters.
The increased number of \ac{BBH} observations coming with the forth LIGO--Virgo--KAGRA observing run will allow us to better measure any correlations between the \ac{BBH} observables, and explore the astrophysics encoded in these correlations.

\section*{acknowledgments}
We thank the groups at Northwestern University for hospitality while this work was developed, in particular to Sylvia Biscoveanu, Monica Gallegos-Garcia, and Miguel Martinez for insightful discussions. 
We thank Alexandra Guerrero for comprehensive review, Daniel Holz for valuable insights, John Veitch for useful technical discussions, \changed{and the anonymous referee for helpful comments}. 
SC is supported by \ac{STFC} studentship 2748218.
CPLB is supported by \ac{STFC} grant ST/V005634/1. 
ZD acknowledges support from the  \ac{CIERA} Board of Visitors Research Professorship. 
SC thanks the \ac{CIERA} Board of Visitors and the University of Glasgow MacRobertson Scholarship for support to visit \ac{CIERA}. 
This research has made use of data from the Gravitational Wave Open Science Center, a service of the LIGO Scientific Collaboration, the Virgo Collaboration, and KAGRA. This material is based upon work supported by the \ac{NSF} LIGO Laboratory which is a major facility fully funded by the \ac{NSF}, as well as \ac{STFC} of the United Kingdom, the Max-Planck-Society (MPS), and the State of Niedersachsen/Germany for support of the construction of Advanced LIGO and construction and operation of the GEO\,600 detector. Additional support for Advanced LIGO was provided by the Australian Research Council. 
Virgo is funded, through the European Gravitational Observatory (EGO), by the French Centre National de Recherche Scientifique (CNRS), the Italian Istituto Nazionale di Fisica Nucleare (INFN) and the Dutch Nikhef, with contributions by institutions from Belgium, Germany, Greece, Hungary, Ireland, Japan, Monaco, Poland, Portugal, Spain. KAGRA is supported by Ministry of Education, Culture, Sports, Science and Technology (MEXT), Japan Society for the Promotion of Science (JSPS) in Japan; National Research Foundation (NRF) and Ministry of Science and ICT (MSIT) in Korea; Academia Sinica (AS) and National Science and Technology Council (NSTC) in Taiwan. 
The authors are grateful for computational resources provided by the LIGO Laboratory and supported by \ac{NSF} Grants PHY-0757058 and PHY-0823459.
This document has been assigned LIGO document number \href{https://dcc.ligo.org/P2500498/public}{LIGO-P2500498}.

These results were produced using models found in \href{https://github.com/scolloms/variable_qmin/tree/v1.0}{this code release}. 
Results and data products used in this work are available on Zenodo: \changed{\dataset[doi:10.5281/zenodo.16895783]{https://doi.org/10.5281/zenodo.16895783}} \citep{variableqminDR}.

\software{\texttt{Bilby} \citep{2019Bilby}, \texttt{gwpopulation} \citep{2019gwpop}, \texttt{gwpopulation\_pipe} \citep{gwpop_pipe}, \texttt{dynesty} \citep{Speagle_2020}, \texttt{NumPy} \citep{harris2020array}, \texttt{matplotlib} \citep{Hunter:2007}, \texttt{corner} \citep{corner}. }

\bibliography{refs}{}

@misc{variableqminDR,
  author = {Colloms, Storm and Doctor, Zoheyr and Berry, Christopher P. L.},
  title = {Data release for Can Big Black Holes Merge with the Smallest Black Holes?},
  url = {https://doi.org/10.5281/zenodo.16895783},
  year = {2025},
}

@article{Talbot:2018cva,
    author = "Talbot, Colm and Thrane, Eric",
    title = "{Measuring the binary black hole mass spectrum with an astrophysically motivated parameterization}",
    eprint = "1801.02699",
    archivePrefix = "arXiv",
    primaryClass = "astro-ph.HE",
    doi = "10.3847/1538-4357/aab34c",
    journal = "Astrophys. J.",
    volume = "856",
    number = "2",
    pages = "173",
    year = "2018"
}

@article{Talbot:2017yur,
    author = "Talbot, Colm and Thrane, Eric",
    title = "{Determining the population properties of spinning black holes}",
    eprint = "1704.08370",
    archivePrefix = "arXiv",
    primaryClass = "astro-ph.HE",
    doi = "10.1103/PhysRevD.96.023012",
    journal = "Phys. Rev. D",
    volume = "96",
    number = "2",
    pages = "023012",
    year = "2017"
}

@article{Fishbach:2025bjh,
    author = "Fishbach, M. and Breivik, K. and Willcox, R. and van Son, L. A. C.",
    title = "{Where are Gaia's small black holes?}",
    eprint = "2508.08986",
    archivePrefix = "arXiv",
    primaryClass = "astro-ph.HE",
    reportNumber = "LIGO-P2500483",
    month = "8",
    year = "2025"
}

@article{Farah:2023swu,
    author = "Farah, Amanda M. and Fishbach, Maya and Holz, Daniel E.",
    title = "{Two of a Kind: Comparing Big and Small Black Holes in Binaries with Gravitational Waves}",
    eprint = "2308.05102",
    archivePrefix = "arXiv",
    primaryClass = "astro-ph.HE",
    doi = "10.3847/1538-4357/ad0558",
    journal = "Astrophys. J.",
    volume = "962",
    number = "1",
    pages = "69",
    year = "2024"
}

@article{LIGOScientific:2024elc,
    author = "Abac, A. G. and others",
    collaboration = "LIGO Scientific, KAGRA, VIRGO",
    title = "{Observation of Gravitational Waves from the Coalescence of a 2.5{\textendash}4.5 M $_{⊙}$ Compact Object and a Neutron Star}",
    eprint = "2404.04248",
    archivePrefix = "arXiv",
    primaryClass = "astro-ph.HE",
    reportNumber = "LIGO-P2300352",
    doi = "10.3847/2041-8213/ad5beb",
    journal = "Astrophys. J. Lett.",
    volume = "970",
    number = "2",
    pages = "L34",
    year = "2024"
}

@article{Fryer:2011cx,
    author = "Fryer, Chris L. and Belczynski, Krzysztof and Wiktorowicz, Grzegorz and Dominik, Michal and Kalogera, Vicky and Holz, Daniel E.",
    title = "{Compact Remnant Mass Function: Dependence on the Explosion Mechanism and Metallicity}",
    eprint = "1110.1726",
    archivePrefix = "arXiv",
    primaryClass = "astro-ph.SR",
    reportNumber = "LA-UR-11-02622",
    doi = "10.1088/0004-637X/749/1/91",
    journal = "Astrophys. J.",
    volume = "749",
    pages = "91",
    year = "2012"
}

@article{Fryer:2022lla,
    author = "Fryer, Chris L. and Olejak, Aleksandra and Belczynski, Krzysztof",
    title = "{The Effect of Supernova Convection On Neutron Star and Black Hole Masses}",
    eprint = "2204.13025",
    archivePrefix = "arXiv",
    primaryClass = "astro-ph.HE",
    reportNumber = "LA-UR-21-32205",
    doi = "10.3847/1538-4357/ac6ac9",
    journal = "Astrophys. J.",
    volume = "931",
    number = "2",
    pages = "94",
    year = "2022"
}

@article{Alvarez-Lopez:2025ltt,
    author = "Alvarez-Lopez, Sofia and Heinzel, Jack and Mould, Matthew and Vitale, Salvatore",
    title = "{Nowhere left to hide: revealing realistic gravitational-wave populations in high dimensions and high resolution with PixelPop}",
    eprint = "2506.20731",
    archivePrefix = "arXiv",
    primaryClass = "astro-ph.HE",
    month = "6",
    year = "2025"
}

@article{Fishbach:2019bbm,
    author = "Fishbach, Maya and Holz, Daniel E.",
    title = "{Picky Partners: The Pairing of Component Masses in Binary Black Hole Mergers}",
    eprint = "1905.12669",
    archivePrefix = "arXiv",
    primaryClass = "astro-ph.HE",
    reportNumber = "LIGO-P1900156",
    doi = "10.3847/2041-8213/ab7247",
    journal = "Astrophys. J. Lett.",
    volume = "891",
    number = "1",
    pages = "L27",
    year = "2020"
}

@article{Farah:2021qom,
    author = "Farah, Amanda M. and Fishbach, Maya and Essick, Reed and Holz, Daniel E. and Galaudage, Shanika",
    title = "{Bridging the Gap: Categorizing Gravitational-wave Events at the Transition between Neutron Stars and Black Holes}",
    eprint = "2111.03498",
    archivePrefix = "arXiv",
    primaryClass = "astro-ph.HE",
    doi = "10.3847/1538-4357/ac5f03",
    journal = "Astrophys. J.",
    volume = "931",
    number = "2",
    pages = "108",
    year = "2022"
}

@article{LIGOScientific:2020kqk,
    author = "Abbott, R. and others",
    collaboration = "LIGO Scientific, Virgo",
    title = "{Population Properties of Compact Objects from the Second LIGO-Virgo Gravitational-Wave Transient Catalog}",
    eprint = "2010.14533",
    archivePrefix = "arXiv",
    primaryClass = "astro-ph.HE",
    reportNumber = "LIGO-P2000077",
    doi = "10.3847/2041-8213/abe949",
    journal = "Astrophys. J. Lett.",
    volume = "913",
    number = "1",
    pages = "L7",
    year = "2021"
}

@article{Essick:2021vlx,
    author = "Essick, Reed and Farah, Amanda and Galaudage, Shanika and Talbot, Colm and Fishbach, Maya and Thrane, Eric and Holz, Daniel E.",
    title = "{Probing Extremal Gravitational-wave Events with Coarse-grained Likelihoods}",
    eprint = "2109.00418",
    archivePrefix = "arXiv",
    primaryClass = "astro-ph.HE",
    doi = "10.3847/1538-4357/ac3978",
    journal = "Astrophys. J.",
    volume = "926",
    number = "1",
    pages = "34",
    year = "2022"
}

@article{Colloms:2025hib,
    author = "Colloms, Storm and Berry, Christopher P. L. and Veitch, John and Zevin, Michael",
    title = "{Exploring the evolution of gravitational-wave emitters with efficient emulation: Constraining the origins of binary black holes using normalising flows}",
    eprint = "2503.03819",
    archivePrefix = "arXiv",
    primaryClass = "astro-ph.HE",
    doi = "10.3847/1538-4357/ade546",
    journal = "Astrophys. J.",
    volume = "988",
    pages = "189",
    year = "2025"
}

@article{LIGOScientific:2020zkf,
    author = "Abbott, R. and others",
    collaboration = "LIGO Scientific, Virgo",
    title = "{GW190814: Gravitational Waves from the Coalescence of a 23 Solar Mass Black Hole with a 2.6 Solar Mass Compact Object}",
    eprint = "2006.12611",
    archivePrefix = "arXiv",
    primaryClass = "astro-ph.HE",
    reportNumber = "LIGO-P190814",
    doi = "10.3847/2041-8213/ab960f",
    journal = "Astrophys. J. Lett.",
    volume = "896",
    number = "2",
    pages = "L44",
    year = "2020"
}

@article{KAGRA:2021duu,
    author = "Abbott, R. and others",
    collaboration = "KAGRA, VIRGO, LIGO Scientific",
    title = "{Population of Merging Compact Binaries Inferred Using Gravitational Waves through GWTC-3}",
    eprint = "2111.03634",
    archivePrefix = "arXiv",
    primaryClass = "astro-ph.HE",
    reportNumber = "LIGO-P2100239 ; Data release: https://zenodo.org/record/5655785, LIGO-P2100239",
    doi = "10.1103/PhysRevX.13.011048",
    journal = "Phys. Rev. X",
    volume = "13",
    number = "1",
    pages = "011048",
    year = "2023"
}

@article{Mapelli:2020vfa,
    author = "Mapelli, Michela",
    title = "{Binary Black Hole Mergers: Formation and Populations}",
    eprint = "2105.12455",
    archivePrefix = "arXiv",
    primaryClass = "astro-ph.HE",
    doi = "10.3389/fspas.2020.00038",
    journal = "Front. Astron. Space Sci.",
    volume = "7",
    pages = "38",
    year = "2020"
}

@article{Mandel:2018hfr,
    author = "Mandel, Ilya and Farmer, Alison",
    title = "{Merging stellar-mass binary black holes}",
    eprint = "1806.05820",
    archivePrefix = "arXiv",
    primaryClass = "astro-ph.HE",
    doi = "10.1016/j.physrep.2022.01.003",
    journal = "Phys. Rept.",
    volume = "955",
    pages = "1--24",
    year = "2022"
}

@article{Olejak:2021fti,
    author = "Olejak, Aleksandra and Belczynski, Krzysztof and Ivanova, Natalia",
    title = "{Impact of common envelope development criteria on the formation of LIGO/Virgo sources}",
    eprint = "2102.05649",
    archivePrefix = "arXiv",
    primaryClass = "astro-ph.HE",
    doi = "10.1051/0004-6361/202140520",
    journal = "Astron. Astrophys.",
    volume = "651",
    pages = "A100",
    year = "2021"
}

@article{vanSon:2021zpk,
    author = "van Son, L. A. C. and de Mink, S. E. and Callister, T. and Justham, S. and Renzo, M. and Wagg, T. and Broekgaarden, F. S. and Kummer, F. and Pakmor, R. and Mandel, I.",
    title = "{The Redshift Evolution of the Binary Black Hole Merger Rate: A Weighty Matter}",
    eprint = "2110.01634",
    archivePrefix = "arXiv",
    primaryClass = "astro-ph.HE",
    doi = "10.3847/1538-4357/ac64a3",
    journal = "Astrophys. J.",
    volume = "931",
    number = "1",
    pages = "17",
    year = "2022"
}

@article{Secunda:2020mhd,
    author = "Secunda, Amy and Bellovary, Jillian and Mac Low, Mordecai-Mark and Ford, K. E. Saavik and McKernan, Barry and Leigh, Nathan W. C. and Lyra, Wladimir and Sandor, Zsolt and Adorno, Jose I.",
    title = "{Orbital Migration of Interacting Stellar Mass Black Holes in Disks around Supermassive Black Holes II. Spins and Incoming Objects}",
    eprint = "2004.11936",
    archivePrefix = "arXiv",
    primaryClass = "astro-ph.HE",
    doi = "10.3847/1538-4357/abbc1d",
    journal = "Astrophys. J.",
    volume = "903",
    number = "2",
    pages = "133",
    year = "2020"
}

@article{Yang:2019okq,
    author = "Yang, Y. and Bartos, I. and Haiman, Z. and Kocsis, B. and Marka, Z. and Stone, N. C. and Marka, S.",
    title = "{AGN Disks Harden the Mass Distribution of Stellar-mass Binary Black Hole Mergers}",
    eprint = "1903.01405",
    archivePrefix = "arXiv",
    primaryClass = "astro-ph.HE",
    doi = "10.3847/1538-4357/ab16e3",
    journal = "Astrophys. J.",
    volume = "876",
    number = "2",
    pages = "122",
    year = "2019"
}

@article{Tagawa:2019osr,
    author = "Tagawa, Hiromichi and Haiman, Zoltan and Kocsis, Bence",
    title = "{Formation and Evolution of Compact Object Binaries in AGN Disks}",
    eprint = "1912.08218",
    archivePrefix = "arXiv",
    primaryClass = "astro-ph.GA",
    doi = "10.3847/1538-4357/ab9b8c",
    journal = "Astrophys. J.",
    volume = "898",
    number = "1",
    pages = "25",
    year = "2020"
}

@article{Bruel:2023bxl,
    author = "Bruel, Tristan and Rodriguez, Carl L. and Lamberts, Astrid and Grudic, Michael Y. and Hafen, Zachary and Feldmann, Robert",
    title = "{Great Balls of FIRE III: Modeling Black Hole Mergers from Massive Star Clusters in Simulations of Galaxies}",
    eprint = "2311.14855",
    archivePrefix = "arXiv",
    primaryClass = "astro-ph.GA",
    doi = "10.1051/0004-6361/202348716",
    journal = "Astron. Astrophys.",
    volume = "686",
    pages = "A106",
    year = "2024"
}

@article{Sedda:2021vjh,
    author = "Sedda, Manuel Arca and Mapelli, Michela and Benacquista, Matthew and Spera, Mario",
    title = "{Isolated and dynamical black hole mergers with B-POP: the role of star formation and dynamics, star cluster evolution, natal kicks, mass and spins, and hierarchical mergers}",
    eprint = "2109.12119",
    archivePrefix = "arXiv",
    primaryClass = "astro-ph.GA",
    doi = "10.1093/mnras/stad331",
    journal = "Mon. Not. Roy. Astron. Soc.",
    volume = "520",
    number = "4",
    pages = "5259--5282",
    year = "2023"
}

@article{Martinez:2021tmr,
    author = "Martinez, Miguel A. S. and Rodriguez, Carl L. and Fragione, Giacomo",
    title = "{On the Mass Ratio Distribution of Black Hole Mergers in Triple Systems}",
    eprint = "2105.01671",
    archivePrefix = "arXiv",
    primaryClass = "astro-ph.SR",
    doi = "10.3847/1538-4357/ac8d55",
    journal = "Astrophys. J.",
    volume = "937",
    number = "2",
    pages = "78",
    year = "2022"
}

@article{Rodriguez:2019huv,
    author = "Rodriguez, Carl L. and Zevin, Michael and Amaro-Seoane, Pau and Chatterjee, Sourav and Kremer, Kyle and Rasio, Frederic A. and Ye, Claire S.",
    title = "{Black holes: The next generation\textemdash{}repeated mergers in dense star clusters and their gravitational-wave properties}",
    eprint = "1906.10260",
    archivePrefix = "arXiv",
    primaryClass = "astro-ph.HE",
    doi = "10.1103/PhysRevD.100.043027",
    journal = "Phys. Rev. D",
    volume = "100",
    number = "4",
    pages = "043027",
    year = "2019"
}

@article{Bailyn:1997xt,
    author = "Bailyn, Charles D. and Jain, Raj K. and Coppi, Paolo and Orosz, Jerome A.",
    title = "{The Mass distribution of stellar black holes}",
    eprint = "astro-ph/9708032",
    archivePrefix = "arXiv",
    doi = "10.1086/305614",
    journal = "Astrophys. J.",
    volume = "499",
    pages = "367",
    year = "1998"
}

@article{Ozel:2010su,
    author = "Ozel, Feryal and Psaltis, Dimitrios and Narayan, Ramesh and McClintock, Jeffrey E.",
    title = "{The Black Hole Mass Distribution in the Galaxy}",
    eprint = "1006.2834",
    archivePrefix = "arXiv",
    primaryClass = "astro-ph.GA",
    doi = "10.1088/0004-637X/725/2/1918",
    journal = "Astrophys. J.",
    volume = "725",
    pages = "1918--1927",
    year = "2010"
}

@article{Farr:2010tu,
    author = "Farr, Will M. and Sravan, Niharika and Cantrell, Andrew and Kreidberg, Laura and Bailyn, Charles D. and Mandel, Ilya and Kalogera, Vicky",
    title = "{The Mass Distribution of Stellar-Mass Black Holes}",
    eprint = "1011.1459",
    archivePrefix = "arXiv",
    primaryClass = "astro-ph.GA",
    doi = "10.1088/0004-637X/741/2/103",
    journal = "Astrophys. J.",
    volume = "741",
    pages = "103",
    year = "2011"
}

@article{Siegel:2022gwc,
    author = "Siegel, Jared C. and others",
    title = "{Investigating the Lower Mass Gap with Low-mass X-Ray Binary Population Synthesis}",
    eprint = "2209.06844",
    archivePrefix = "arXiv",
    primaryClass = "astro-ph.HE",
    doi = "10.3847/1538-4357/ace9d9",
    journal = "Astrophys. J.",
    volume = "954",
    number = "2",
    pages = "212",
    year = "2023"
}

@article{Moe_2017,
   title={Mind Your Ps and Qs: The Interrelation between Period (P) and Mass-ratio (Q) Distributions of Binary Stars},
   volume={230},
   ISSN={1538-4365},
   url={http://dx.doi.org/10.3847/1538-4365/aa6fb6},
   DOI={10.3847/1538-4365/aa6fb6},
   number={2},
   journal={Astrophys. J. Suppl.},
   publisher={American Astronomical Society},
   author={Moe, Maxwell and Di Stefano, Rosanne},
   year={2017},
   month=jun, pages={15} }

@article{Thompson:2018ycv,
    author = "Thompson, Todd A. and others",
        title = "{A noninteracting low-mass black hole-giant star binary system}",
      journal = {Science},
         year = 2019,
        month = nov,
       volume = {366},
       number = {6465},
        pages = {637-640},
    eprint = "1806.02751",
    archivePrefix = "arXiv",
    primaryClass = "astro-ph.HE",
    doi = "10.1126/science.aau4005",
    month = "6",
    year = "2018"
}

@article{Ray:2025aqr,
    author = "Ray, Anarya and Farr, Will and Kalogera, Vicky",
    title = "{Hiding Out at the Low End: No Gap and a Peak in the Black-Hole Mass Spectrum}",
    eprint = "2507.09099",
    archivePrefix = "arXiv",
    primaryClass = "astro-ph.HE",
    month = "7",
    year = "2025"
}

@article{Patton:2021gwh,
    author = "Patton, Rachel A. and Sukhbold, Tuguldur and Eldridge, J. J.",
    title = "{Comparing compact object distributions from mass- and presupernova core structure-based prescriptions}",
    eprint = "2106.05978",
    archivePrefix = "arXiv",
    primaryClass = "astro-ph.HE",
    doi = "10.1093/mnras/stab3797",
    journal = "Mon. Not. Roy. Astron. Soc.",
    volume = "511",
    number = "1",
    pages = "903--913",
    year = "2022"
}

@article{Liu:2020uba,
    author = "Liu, Tong and Wei, Yun-Feng and Xue, Li and Sun, Mou-Yuan",
    title = "{Final Compact Remnants in Core-collapse Supernovae from 20 to 40 $M_\odot$: The Lower Mass Gap}",
    eprint = "2011.14361",
    archivePrefix = "arXiv",
    primaryClass = "astro-ph.HE",
    doi = "10.3847/1538-4357/abd24e",
    journal = "Astrophys. J.",
    volume = "908",
    number = "1",
    pages = "106",
    year = "2021"
}

@article{Belczynski:2011bn,
    author = "Belczynski, K. and Wiktorowicz, G. and Fryer, C. and Holz, D. and Kalogera, V.",
    title = "{Missing Black Holes Unveil The Supernova Explosion Mechanism}",
    eprint = "1110.1635",
    archivePrefix = "arXiv",
    primaryClass = "astro-ph.GA",
    doi = "10.1088/0004-637X/757/1/91",
    journal = "Astrophys. J.",
    volume = "757",
    pages = "91",
    year = "2012"
}

@article{Jayasinghe:2021uqb,
    author = "Jayasinghe, T. and others",
    title = "{A unicorn in monoceros: the 3{\,}M{\ensuremath{\odot}} dark companion to the bright, nearby red giant V723 Mon is a non-interacting, mass-gap black hole candidate}",
    eprint = "2101.02212",
    archivePrefix = "arXiv",
    primaryClass = "astro-ph.SR",
    doi = "10.1093/mnras/stab907",
    journal = "Mon. Not. Roy. Astron. Soc.",
    volume = "504",
    number = "2",
    pages = "2577--2602",
    year = "2021"
}

@article{Speagle_2020,
   title={dynesty: a dynamic nested sampling package for estimating Bayesian posteriors and evidences},
   volume={493},
   ISSN={1365-2966},
   url={http://dx.doi.org/10.1093/mnras/staa278},
   DOI={10.1093/mnras/staa278},
   number={3},
   journal={Mon. Not. Roy. Astron. Soc.},
   publisher={Oxford University Press (OUP)},
   author={Speagle, Joshua S},
   year={2020},
   month=feb, pages={3132–3158} }

@inbook{Mapelli:2021taw,
    author = "Mapelli, Michela",
    title = "{Formation Channels of Single and Binary Stellar-Mass Black Holes}",
    eprint = "2106.00699",
    archivePrefix = "arXiv",
    primaryClass = "astro-ph.HE",
    doi = "10.1007/978-981-15-4702-7_16-1",
    year = "2021",
    booktitle = {Handbook of Gravitational Wave Astronomy}
}

@article{Callister:2024cdx,
    author = "Callister, T. A.",
    title = "{Observed Gravitational-Wave Populations}",
    eprint = "2410.19145",
    archivePrefix = "arXiv",
    primaryClass = "astro-ph.HE",
    month = "10",
    year = "2024"
}

@article{Galaudage:2024meo,
    author = "Galaudage, Shanika and Lamberts, Astrid",
    title = "{Compactness peaks: An astrophysical interpretation of the mass distribution of merging binary black holes}",
    eprint = "2407.17561",
    archivePrefix = "arXiv",
    primaryClass = "astro-ph.HE",
    doi = "10.1051/0004-6361/202451654",
    journal = "Astron. Astrophys.",
    volume = "694",
    pages = "A186",
    year = "2025"
}

@ARTICLE{2019Bilby,
       author = {{Ashton}, Gregory and {H{\"u}bner}, Moritz and {Lasky}, Paul D. and {Talbot}, Colm and {Ackley}, Kendall and {Biscoveanu}, Sylvia and {Chu}, Qi and {Divakarla}, Atul and {Easter}, Paul J. and {Goncharov}, Boris and {Hernandez Vivanco}, Francisco and {Harms}, Jan and {Lower}, Marcus E. and {Meadors}, Grant D. and {Melchor}, Denyz and {Payne}, Ethan and {Pitkin}, Matthew D. and {Powell}, Jade and {Sarin}, Nikhil and {Smith}, Rory J.~E. and {Thrane}, Eric},
        title = "{BILBY: A User-friendly Bayesian Inference Library for Gravitational-wave Astronomy}",
      journal = {\apjs},
         year = 2019,
        month = apr,
       volume = {241},
       number = {2},
          eid = {27},
        pages = {27},
          doi = {10.3847/1538-4365/ab06fc},
archivePrefix = {arXiv},
       eprint = {1811.02042},
 primaryClass = {astro-ph.IM},
       adsurl = {https://ui.adsabs.harvard.edu/abs/2019ApJS..241...27A}
}

@ARTICLE{2019gwpop,
       author = {{Talbot}, Colm and {Smith}, Rory and {Thrane}, Eric and {Poole}, Gregory B.},
        title = "{Parallelized inference for gravitational-wave astronomy}",
      journal = {"Phys. Rev. D},
         year = 2019,
        month = aug,
       volume = {100},
       number = {4},
          eid = {043030},
        pages = {043030},
          doi = {10.1103/PhysRevD.100.043030},
archivePrefix = {arXiv},
       eprint = {1904.02863},
 primaryClass = {astro-ph.IM},
       adsurl = {https://ui.adsabs.harvard.edu/abs/2019PhRvD.100d3030T}
}

@ARTICLE{GWTC3_sensitivity,
    title={GWTC-3: Compact Binary Coalescences Observed by LIGO and Virgo During the Second Part of the Third Observing Run — O1+O2+O3 Search Sensitivity Estimates},
    DOI={10.5281/zenodo.5636815},
    publisher={Zenodo},
    author={Abbott, R. and others},
    collaboration={LIGO Scientific Collaboration and Virgo Collaboration and KAGRA Collaboration},
    year={2023},
    month={May},
}

@ARTICLE{gwpop_pipe,
    title={GWPopulation pipe},
    DOI={10.5281/zenodo.5654673},
    publisher={Zenodo},
    author={Talbot, Colm},
    year={2021},
    month={Nov},
    url={https://git.ligo.org/RatesAndPopulations/gwpopulation_pipe}
}

@article{Talbot:2024yqw,
    author = "Talbot, Colm and Farah, Amanda and Galaudage, Shanika and Golomb, Jacob and Tong, Hui",
    title = "{GWPopulation: Hardware agnostic population inference for compact binaries and beyond}",
    eprint = "2409.14143",
    archivePrefix = "arXiv",
    primaryClass = "astro-ph.IM",
    doi = "10.21105/joss.07753",
    journal = "J. Open Source Softw.",
    volume = "10",
    number = "109",
    pages = "7753",
    year = "2025"
}

@article{Safarzadeh:2020mlb,
    author = "Safarzadeh, Mohammadtaher and Farr, Will M. and Ramirez-Ruiz, Enrico",
    title = "{A trend in the effective spin distribution of LIGO binary black holes with mass}",
    eprint = "2001.06490",
    archivePrefix = "arXiv",
    primaryClass = "gr-qc",
    doi = "10.3847/1538-4357/ab80be",
    journal = "Astrophys. J.",
    volume = "894",
    number = "2",
    pages = "129",
    year = "2020"
}

@article{Ray:2024hos,
    author = "Ray, Anarya and Maga{\~n}a Hernandez, Ignacio and Breivik, Katelyn and Creighton, Jolien",
    title = "{Searching for binary black hole sub-populations in gravitational wave data using binned Gaussian processes}",
    eprint = "2404.03166",
    archivePrefix = "arXiv",
    primaryClass = "astro-ph.HE",
    reportNumber = "LIGO-P2400115",
    month = "4",
    year = "2024"
}

@article{corner,
      doi = {10.21105/joss.00024},
      url = {https://doi.org/10.21105/joss.00024},
      year  = {2016},
      month = {jun},
      publisher = {The Open Journal},
      volume = {1},
      number = {2},
      pages = {24},
      author = {Daniel Foreman-Mackey},
      title = {corner.py: Scatterplot matrices in Python},
      journal = {J. Open Source Softw.}
    }

@Article{Hunter:2007,
  Author    = {Hunter, J. D.},
  Title     = {Matplotlib: A 2D graphics environment},
  Journal   = {	Comput. Sci. Eng.},
  Volume    = {9},
  Number    = {3},
  Pages     = {90--95},
  publisher = {IEEE COMPUTER SOC},
  doi       = {10.1109/MCSE.2007.55},
  year      = 2007
}

@Article{harris2020array,
 title         = {Array programming with {NumPy}},
 author        = {Charles R. Harris and K. Jarrod Millman and St{\'{e}}fan J.
                 van der Walt and Ralf Gommers and Pauli Virtanen and David
                 Cournapeau and Eric Wieser and Julian Taylor and Sebastian
                 Berg and Nathaniel J. Smith and Robert Kern and Matti Picus
                 and Stephan Hoyer and Marten H. van Kerkwijk and Matthew
                 Brett and Allan Haldane and Jaime Fern{\'{a}}ndez del
                 R{\'{i}}o and Mark Wiebe and Pearu Peterson and Pierre
                 G{\'{e}}rard-Marchant and Kevin Sheppard and Tyler Reddy and
                 Warren Weckesser and Hameer Abbasi and Christoph Gohlke and
                 Travis E. Oliphant},
 year          = {2020},
 month         = sep,
 journal       = {Nature},
 volume        = {585},
 number        = {7825},
 pages         = {357--362},
 doi           = {10.1038/s41586-020-2649-2},
 publisher     = {Springer Science and Business Media {LLC}},
 url           = {https://doi.org/10.1038/s41586-020-2649-2}
}

@article{Fishbach:2018edt,
    author = "Fishbach, Maya and Holz, Daniel E. and Farr, Will M.",
    title = "{Does the Black Hole Merger Rate Evolve with Redshift?}",
    eprint = "1805.10270",
    archivePrefix = "arXiv",
    primaryClass = "astro-ph.HE",
    doi = "10.3847/2041-8213/aad800",
    journal = "Astrophys. J. Lett.",
    volume = "863",
    number = "2",
    pages = "L41",
    year = "2018"
}

@article{brielUnderstandingHighmassBinary2022,
  title = {Understanding the High-Mass Binary Black Hole Population from Stable Mass Transfer and Super-{{Eddington}} Accretion in {{BPASS}}},
  author = {Briel, M. M. and Stevance, H. F. and Eldridge, J. J.},
  year = {2022},
  eprint = {2206.13842},
  eprinttype = {arxiv},
  eprintclass = {astro-ph},
  doi = "10.1093/mnras/stad399",
  journal = "Mon. Not. Roy. Astron. Soc.",
  urldate = {2022-11-29},
  langid = {english},
    volume = "520",
    number = "4",
    pages = "5724--5745"
}

@article{Adamcewicz:2022hce,
    author = "Adamcewicz, Christian and Thrane, Eric",
    title = "{Do unequal-mass binary black hole systems have larger \ensuremath{\chi}eff? Probing correlations with copulas in gravitational-wave astronomy}",
    eprint = "2208.03405",
    archivePrefix = "arXiv",
    primaryClass = "astro-ph.HE",
    doi = "10.1093/mnras/stac2961",
    journal = "Mon. Not. Roy. Astron. Soc.",
    volume = "517",
    number = "3",
    pages = "3928--3937",
    year = "2022"
}

@article{Li:2022jge,
    author = "Li, Yin-Jie and Wang, Yuan-Zhu and Tang, Shao-Peng and Yuan, Qiang and Fan, Yi-Zhong and Wei, Da-Ming",
    title = "{Divergence in Mass Ratio Distributions between Low-mass and High-mass Coalescing Binary Black Holes}",
    eprint = "2201.01905",
    archivePrefix = "arXiv",
    primaryClass = "astro-ph.HE",
    doi = "10.3847/2041-8213/ac78dd",
    journal = "Astrophys. J. Lett.",
    volume = "933",
    number = "1",
    pages = "L14",
    year = "2022"
}

@article{Li:2024jzi,
    author = "Li, Yin-Jie and Tang, Shao-Peng and Gao, Shi-Jie and Wu, Dao-Cheng and Wang, Yuan-Zhu",
    title = "{Exploring Field-evolution and Dynamical-capture Coalescing Binary Black Holes in GWTC-3}",
    eprint = "2404.09668",
    archivePrefix = "arXiv",
    primaryClass = "astro-ph.HE",
    doi = "10.3847/1538-4357/ad83b5",
    journal = "Astrophys. J.",
    volume = "977",
    number = "1",
    pages = "67",
    year = "2024"
}

@article{Rinaldi:2025emt,
    author = "Rinaldi, Stefano and Liang, Yajie and Demasi, Gabriele and Mapelli, Michela and Del Pozzo, Walter",
    title = "{Exploration of features in the black hole mass spectrum inspired by non-parametric analyses of gravitational wave observations}",
    eprint = "2506.05929",
    archivePrefix = "arXiv",
    primaryClass = "astro-ph.HE",
    month = "6",
    year = "2025"
}

@article{Roy:2025ktr,
    author = "Roy, Soumendra Kishore and van Son, Lieke A. C. and Farr, Will M.",
    title = "{A Mid-Thirties Crisis: Dissecting the Properties of Gravitational Wave Sources Near the 35 Solar Mass Peak}",
    eprint = "2507.01086",
    archivePrefix = "arXiv",
    primaryClass = "astro-ph.HE",
    reportNumber = "LIGO document number LIGO-P2500403",
    month = "7",
    year = "2025"
}

@article{LIGOScientific:2014pky,
    author = "Aasi, J. and others",
    collaboration = "LIGO Scientific",
    title = "{Advanced LIGO}",
    eprint = "1411.4547",
    archivePrefix = "arXiv",
    primaryClass = "gr-qc",
    doi = "10.1088/0264-9381/32/7/074001",
    journal = "Class. Quant. Grav.",
    volume = "32",
    pages = "074001",
    year = "2015"
}

@article{VIRGO:2014yos,
    author = "Acernese, F. and others",
    collaboration = "Virgo",
    title = "{Advanced Virgo: a second-generation interferometric gravitational wave detector}",
    eprint = "1408.3978",
    archivePrefix = "arXiv",
    primaryClass = "gr-qc",
    doi = "10.1088/0264-9381/32/2/024001",
    journal = "Class. Quant. Grav.",
    volume = "32",
    number = "2",
    pages = "024001",
    year = "2015"
}

@article{KAGRA:2020tym,
    author = "Akutsu, T. and others",
    collaboration = "KAGRA",
    title = "{Overview of KAGRA: Detector design and construction history}",
    eprint = "2005.05574",
    archivePrefix = "arXiv",
    primaryClass = "physics.ins-det",
    doi = "10.1093/ptep/ptaa125",
    journal = "PTEP",
    volume = "2021",
    number = "5",
    pages = "05A101",
    year = "2021"
}

@article{Callister:2023tgi,
    author = "Callister, Thomas A. and Farr, Will M.",
    title = "{Parameter-Free Tour of the Binary Black Hole Population}",
    eprint = "2302.07289",
    archivePrefix = "arXiv",
    primaryClass = "astro-ph.HE",
    doi = "10.1103/PhysRevX.14.021005",
    journal = "Phys. Rev. X",
    volume = "14",
    number = "2",
    pages = "021005",
    year = "2024"
}

@article{Rinaldi:2023bbd,
    author = "Rinaldi, Stefano and Del Pozzo, Walter and Mapelli, Michela and Lorenzo-Medina, Ana and Dent, Thomas",
    title = "{Evidence of evolution of the black hole mass function with redshift}",
    eprint = "2310.03074",
    archivePrefix = "arXiv",
    primaryClass = "astro-ph.HE",
    doi = "10.1051/0004-6361/202348161",
    journal = "Astron. Astrophys.",
    volume = "684",
    pages = "A204",
    year = "2024"
}

@article{Talbot:2023pex,
    author = "Talbot, Colm and Golomb, Jacob",
    title = "{Growing pains: understanding the impact of likelihood uncertainty on hierarchical Bayesian inference for gravitational-wave astronomy}",
    eprint = "2304.06138",
    archivePrefix = "arXiv",
    primaryClass = "astro-ph.IM",
    doi = "10.1093/mnras/stad2968",
    journal = "Mon. Not. Roy. Astron. Soc.",
    volume = "526",
    number = "3",
    pages = "3495--3503",
    year = "2023"
}

@article{fishbachWhereAreLIGO2017,
  author = "Fishbach, Maya and Holz, Daniel E.",
    title = "{Where Are LIGO{\textquoteright}s Big Black Holes?}",
    eprint = "1709.08584",
    archivePrefix = "arXiv",
    primaryClass = "astro-ph.HE",
    doi = "10.3847/2041-8213/aa9bf6",
    journal = "Astrophys. J. Lett.",
    volume = "851",
    number = "2",
    pages = "L25",
    year = "2017"
}

@article{Kovetz:2016kpi,
    author = "Kovetz, Ely D. and Cholis, Ilias and Breysse, Patrick C. and Kamionkowski, Marc",
    title = "{Black hole mass function from gravitational wave measurements}",
    eprint = "1611.01157",
    archivePrefix = "arXiv",
    primaryClass = "astro-ph.CO",
    doi = "10.1103/PhysRevD.95.103010",
    journal = "Phys. Rev. D",
    volume = "95",
    number = "10",
    pages = "103010",
    year = "2017"
}

@article{Zevin:2020gma,
    author = "Zevin, Michael and Spera, Mario and Berry, Christopher P. L. and Kalogera, Vicky",
    title = "{Exploring the Lower Mass Gap and Unequal Mass Regime in Compact Binary Evolution}",
    eprint = "2006.14573",
    archivePrefix = "arXiv",
    primaryClass = "astro-ph.HE",
    doi = "10.3847/2041-8213/aba74e",
    journal = "Astrophys. J. Lett.",
    volume = "899",
    pages = "L1",
    year = "2020"
}

@article{godfreyCosmicCousinsIdentification2023,
  title = {Cosmic {{Cousins}}: {{Identification}} of a {{Subpopulation}} of {{Binary Black Holes Consistent}} with {{Isolated Binary Evolution}}},
  shorttitle = {Cosmic {{Cousins}}},
  author = {Godfrey, Jaxen and Edelman, Bruce and Farr, Ben},
  year = {2023},
  url = {http://arxiv.org/abs/2304.01288},
  urldate = {2023-04-20},
  langid = {english},
  pubstate = {preprint},
archivePrefix = {arXiv},
       eprint = {2304.01288},
 primaryClass = {astro-ph.HE},
}

@article{Heinzel:2023hlb,
    author = "Heinzel, Jack and Biscoveanu, Sylvia and Vitale, Salvatore",
    title = "{Probing Correlations in the Binary Black Hole Population with Flexible Models}",
    eprint = "2312.00993",
    archivePrefix = "arXiv",
    primaryClass = "astro-ph.HE",
    month = "12",
    doi = "10.1103/PhysRevD.109.103006",
    journal = "Phys. Rev. D",
    volume = "109",
    number = "10",
    pages = "103006",
    year = "2024"
}

@article{Sadiq:2023zee,
    author = "Sadiq, Jam and Dent, Thomas and Gieles, Mark",
    title = "{Binary Vision: The Mass Distribution of Merging Binary Black Holes via Iterative Density Estimation}",
    eprint = "2307.12092",
    archivePrefix = "arXiv",
    primaryClass = "astro-ph.HE",
    doi = "10.3847/1538-4357/ad0ce6",
    journal = "Astrophys. J.",
    volume = "960",
    number = "1",
    pages = "65",
    year = "2024"
}

@article{mandelExtractingDistributionParameters2019,
  title = {Extracting Distribution Parameters from Multiple Uncertain Observations with Selection Biases},
  author = {Mandel, Ilya and Farr, Will M and Gair, Jonathan R},
  date = {2019-06-11},
  year = {2019},
  journal = {Mon. Not. Roy. Astron. Soc.},
  volume = {486},
  number = {1},
  pages = {1086--1093},
  issn = {0035-8711, 1365-2966},
  doi = {10.1093/mnras/stz896},
  url = {https://academic.oup.com/mnras/article/486/1/1086/5421632},
  urldate = {2022-10-17},
  langid = {english}
}

@article{thraneIntroductionBayesianInference2019,
  title = {An Introduction to {{Bayesian}} Inference in Gravitational-Wave Astronomy: Parameter Estimation, Model Selection, and Hierarchical Models},
  shorttitle = {An Introduction to {{Bayesian}} Inference in Gravitational-Wave Astronomy},
  author = {Thrane, Eric and Talbot, Colm},
  year = {2019},
  journal = {Publications of the Astronomical Society of Australia},
  shortjournal = {Publ. Astron. Soc. Aust.},
  volume = {36},
  eprint = {1809.02293},
  eprinttype = {arxiv},
  eprintclass = {astro-ph},
  pages = {e010},
  issn = {1323-3580, 1448-6083},
  doi = {10.1017/pasa.2019.2},
  url = {http://arxiv.org/abs/1809.02293},
  urldate = {2022-10-21},
  langid = {english}
}

@incollection{vitaleInferringPropertiesPopulation2021,
  address = {Singapore},
	author = {Vitale, Salvatore and Gerosa, Davide and Farr, Will M. and Taylor, Stephen R.},
	booktitle = {Handbook of Gravitational Wave Astronomy},
	doi = {10.1007/978-981-15-4702-7_45-1},
	editor = {Bambi, Cosimo and Katsanevas, Stavros and Kokkotas, Konstantinos D.},
	isbn = {978-981-15-4702-7},
	pages = {1--60},
	publisher = {Springer Singapore},
	title = {Inferring the Properties of a Population of Compact Binaries in Presence of Selection Effects},
	url = {https://doi.org/10.1007/978-981-15-4702-7_45-1},
	year = {2020}
}

@article{LVKsensitivity2018,
	author = {Abbott, B. P. and Abbott, R. and Abbott, T. D. and Abernathy, M. R. and Acernese, F. and Ackley, K. and others},
	date = {2018/04/26},
	doi = {10.1007/s41114-018-0012-9},
	id = {Abbott2018},
	isbn = {1433-8351},
	journal = {Living Rev. Rel.},
	number = {1},
	pages = {3},
	title = {Prospects for observing and localizing gravitational-wave transients with Advanced LIGO, Advanced Virgo and KAGRA},
	url = {https://doi.org/10.1007/s41114-018-0012-9},
	volume = {21},
	year = {2018}}

@article{Li:2023yyt,
    author = "Li, Yin-Jie and Wang, Yuan-Zhu and Tang, Shao-Peng and Fan, Yi-Zhong",
    title = "{Resolving the Stellar-Collapse and Hierarchical-Merger Origins of the Coalescing Black Holes}",
    eprint = "2303.02973",
    archivePrefix = "arXiv",
    primaryClass = "astro-ph.HE",
    doi = "10.1103/PhysRevLett.133.051401",
    journal = "Phys. Rev. Lett.",
    volume = "133",
    number = "5",
    pages = "051401",
    year = "2024"
}

@article{Koehn:2024set,
    author = "Koehn, Hauke and others",
    title = "{From existing and new nuclear and astrophysical constraints to stringent limits on the equation of state of neutron-rich dense matter}",
    eprint = "2402.04172",
    archivePrefix = "arXiv",
    primaryClass = "astro-ph.HE",
    reportNumber = "LA-UR-24-20420",
    doi = "10.1103/PhysRevX.15.021014",
    journal = "Phys. Rev. X",
    volume = "15",
    number = "2",
    pages = "021014",
    year = "2025"
}

@article{Rutherford:2024srk,
    author = "Rutherford, Nathan and others",
    title = "{Constraining the Dense Matter Equation of State with New NICER Mass{\textendash}Radius Measurements and New Chiral Effective Field Theory Inputs}",
    eprint = "2407.06790",
    archivePrefix = "arXiv",
    primaryClass = "astro-ph.HE",
    doi = "10.3847/2041-8213/ad5f02",
    journal = "Astrophys. J. Lett.",
    volume = "971",
    number = "1",
    pages = "L19",
    year = "2024"
}

@article{Antonini:2024het,
    author = "Antonini, Fabio and Romero-Shaw, Isobel M. and Callister, Thomas",
    title = "{Star Cluster Population of High Mass Black Hole Mergers in Gravitational Wave Data}",
    eprint = "2406.19044",
    archivePrefix = "arXiv",
    primaryClass = "astro-ph.HE",
    doi = "10.1103/PhysRevLett.134.011401",
    journal = "Phys. Rev. Lett.",
    volume = "134",
    number = "1",
    pages = "011401",
    year = "2025"
}

@article{Baibhav:2022qxm,
    author = "Baibhav, Vishal and Doctor, Zoheyr and Kalogera, Vicky",
    title = "{Dropping Anchor: Understanding the Populations of Binary Black Holes with Random and Aligned-spin Orientations}",
    eprint = "2212.12113",
    archivePrefix = "arXiv",
    primaryClass = "astro-ph.HE",
    doi = "10.3847/1538-4357/acbf4c",
    journal = "Astrophys. J.",
    volume = "946",
    number = "1",
    pages = "50",
    year = "2023"
}

@article{zevinOneChannelRule2021,
  title = {One {{Channel}} to {{Rule Them All}}? {{Constraining}} the {{Origins}} of {{Binary Black Holes}} Using {{Multiple Formation Pathways}}},
  shorttitle = {One {{Channel}} to {{Rule Them All}}?},
  author = {Zevin, Michael and Bavera, Simone S. and Berry, Christopher P. L. and Kalogera, Vicky and Fragos, Tassos and Marchant, Pablo and Rodriguez, Carl L. and Antonini, Fabio and Holz, Daniel E. and Pankow, Chris},
  date = {2021-04-01},
  year={2021},
  journal = {Astrophys. J.},
  shortjournal = {ApJ},
  volume = {910},
  number = {2},
  eprint = {2011.10057},
  eprinttype = {arxiv},
  eprintclass = {astro-ph, physics:gr-qc},
  pages = {152},
  issn = {0004-637X, 1538-4357},
  doi = {10.3847/1538-4357/abe40e},
  url = {http://arxiv.org/abs/2011.10057},
  urldate = {2022-10-04}
}

@article{LIGOScientific:2021djp,
    author = "Abbott, R. and others",
    collaboration = "LIGO Scientific, Virgo, KAGRA",
    title = "{GWTC-3: Compact Binary Coalescences Observed by LIGO and Virgo during the Second Part of the Third Observing Run}",
    eprint = "2111.03606",
    archivePrefix = "arXiv",
    primaryClass = "gr-qc",
    reportNumber = "LIGO-P2000318",
    doi = "10.1103/PhysRevX.13.041039",
    journal = "Phys. Rev. X",
    volume = "13",
    number = "4",
    pages = "041039",
    year = "2023"
}

@article{LIGOScientific:2021usb,
    author = "Abbott, R. and others",
    collaboration = "LIGO Scientific, VIRGO",
    title = "{GWTC-2.1: Deep extended catalog of compact binary coalescences observed by LIGO and Virgo during the first half of the third observing run}",
    eprint = "2108.01045",
    archivePrefix = "arXiv",
    primaryClass = "gr-qc",
    reportNumber = "LIGO-P2100063",
    doi = "10.1103/PhysRevD.109.022001",
    journal = "Phys. Rev. D",
    volume = "109",
    number = "2",
    pages = "022001",
    year = "2024"
}

@article{Mandel:2021smh,
    author = "Mandel, Ilya and Broekgaarden, Floor S.",
    title = "{Rates of compact object coalescences}",
    eprint = "2107.14239",
    archivePrefix = "arXiv",
    primaryClass = "astro-ph.HE",
    doi = "10.1007/s41114-021-00034-3",
    journal = "Living Rev. Rel.",
    volume = "25",
    number = "1",
    pages = "1",
    year = "2022"
}

@article{Belczynski:2021zaz,
    author = "Belczynski, K. and Romagnolo, A. and Olejak, A. and Klencki, J. and Chattopadhyay, D. and Stevenson, S. and Miller, M. Coleman and Lasota, J. -P. and Crowther, Paul A.",
    title = "{The Uncertain Future of Massive Binaries Obscures the Origin of LIGO/Virgo Sources}",
    eprint = "2108.10885",
    archivePrefix = "arXiv",
    primaryClass = "astro-ph.HE",
    doi = "10.3847/1538-4357/ac375a",
    journal = "Astrophys. J.",
    volume = "925",
    number = "1",
    pages = "69",
    year = "2022"
}

@article{Sedda:2020vwo,
    author = "Sedda, Manuel Arca and Mapelli, Michela and Spera, Mario and Benacquista, Matthew and Giacobbo, Nicola",
    title = "{Fingerprints of binary black hole formation channels encoded in the mass and spin of merger remnants}",
    eprint = "2003.07409",
    archivePrefix = "arXiv",
    primaryClass = "astro-ph.GA",
    doi = "10.3847/1538-4357/ab88b2",
    journal = "Astrophys. J.",
    volume = "894",
    number = "2",
    pages = "133",
    year = "2020"
}

@article{Wysocki:2018mpo,
    author = "Wysocki, Daniel and Lange, Jacob and O'Shaughnessy, Richard",
    title = "{Reconstructing phenomenological distributions of compact binaries via gravitational wave observations}",
    eprint = "1805.06442",
    archivePrefix = "arXiv",
    primaryClass = "gr-qc",
    reportNumber = "LIGO P1800107, LIGO-P1800107",
    doi = "10.1103/PhysRevD.100.043012",
    journal = "Phys. Rev. D",
    volume = "100",
    number = "4",
    pages = "043012",
    year = "2019"
}

@article{neijsselEffectMetallicityspecificStar2019,
  title = {The Effect of the Metallicity-Specific Star Formation History on Double Compact Object Mergers},
  author = {Neijssel, Coenraad J and Vigna-Gómez, Alejandro and Stevenson, Simon and Barrett, Jim W and Gaebel, Sebastian M and Broekgaarden, Floor S and {de~Mink}, Selma E and Szécsi, Dorottya and Vinciguerra, Serena and Mandel, Ilya},
  year = {2019},
  journal = {Mon. Not. Roy. Astron. Soc.},
  volume = {490},
  number = {3},
  pages = {3740--3759},
  issn = {0035-8711, 1365-2966},
  doi = {10.1093/mnras/stz2840},
  url = {https://academic.oup.com/mnras/article/490/3/3740/5588621},
  urldate = {2022-11-29},
  langid = {english}
}

@article{Broekgaarden:2022nst,
    author = "Broekgaarden, Floor S. and Stevenson, Simon and Thrane, Eric",
    title = "{Signatures of Mass Ratio Reversal in Gravitational Waves from Merging Binary Black Holes}",
    eprint = "2205.01693",
    archivePrefix = "arXiv",
    primaryClass = "astro-ph.HE",
    doi = "10.3847/1538-4357/ac8879",
    journal = "Astrophys. J.",
    volume = "938",
    number = "1",
    pages = "45",
    year = "2022"
}

@ARTICLE{Bate8134774,
  author={Bate, Matthew R. and Bonnell, Ian A.},
  journal={Mon. Not. Roy. Astron. Soc.}, 
  title={Accretion during binary star formation — II. Gaseous accretion and disc formation}, 
  year={1997},
  volume={285},
  number={1},
  pages={33-48},
  doi={10.1093/mnras/285.1.33}}

@article{Cheng:2023ddt,
    author = "Cheng, April Qiu and Zevin, Michael and Vitale, Salvatore",
    title = "{What You Don\textquoteright{}t Know Can Hurt You: Use and Abuse of Astrophysical Models in Gravitational-wave Population Analyses}",
    eprint = "2307.03129",
    archivePrefix = "arXiv",
    primaryClass = "astro-ph.HE",
    reportNumber = "LIGO DCC: P2300200",
    doi = "10.3847/1538-4357/aced98",
    journal = "Astrophys. J.",
    volume = "955",
    number = "2",
    pages = "127",
    year = "2023"
}
\bibliographystyle{aasjournal}

\end{document}